\documentclass[fleqn,usenatbib]{mnras}

\usepackage[T1]{fontenc}

\DeclareRobustCommand{\VAN}[3]{#2}
\let\VANthebibliography\thebibliography
\def\thebibliography{\DeclareRobustCommand{\VAN}[3]{##3}\VANthebibliography}

\usepackage{graphicx}
\usepackage{enumerate}
\usepackage{newtxtext,newtxmath}
\usepackage[T1]{fontenc}
\usepackage{gensymb}
\usepackage{mathtools}
\usepackage{physics}
\usepackage{natbib}
\usepackage{hyperref}
\usepackage{color}
\usepackage{ulem}
\usepackage{soul}
\usepackage{url}
\usepackage{xspace}
\usepackage{multirow}
\usepackage{pdflscape}

\newcommand{\kms}{\mbox{km\,s$^{-1}$}}

\defcitealias{GRAVITY_2018}{GC18}

\title[WISDOM: Metrics for BH mass measurements]{WISDOM Project -- XIX.\ Figures of merit for supermassive black hole mass measurements using molecular gas and/or megamaser kinematics}

\author[H.\ Zhang et al.]{
Hengyue Zhang,$^{1}$\thanks{E-mail: hengyue.zhang@physics.ox.ac.uk}
Martin Bureau,$^{1}$\thanks{E-mail: martin.bureau@physics.ox.ac.uk}
Mark D.\ Smith,$^{1}$
Michele Cappellari,$^{1}$
Timothy A.\ Davis,$^{2}$
\newauthor{
Pandora Dominiak,$^{1}$
Jacob S.\ Elford,$^{2}$
Fu-Heng Liang,$^{1}$
Ilaria Ruffa$^{2,3}$} and
Thomas G.\ Williams$^{1}$
\\
$^{1}$Sub-department of Astrophysics, Department of Physics, University of Oxford, Denys Wilkinson Building, Keble Road, Oxford, OX1~3RH, UK\\
$^{2}$School of Physics \& Astronomy, Cardiff University, Queens Buildings, The Parade, Cardiff, CF24~3AA, UK\\
$^{3}$INAF - Istituto di Radioastronomia, via P.\ Gobetti 101, 40129 Bologna, Italy \\
}

\date{Accepted XXX. Received YYY; in original form ZZZ}
\pubyear{2024}

\begin{document}
\label{firstpage}
\pagerange{\pageref{firstpage}--\pageref{lastpage}}
    \maketitle

\begin{abstract}
  The mass ($M_\mathrm{BH}$) of a supermassive black hole (SMBH) can be measured using spatially-resolved kinematics of the region where the SMBH dominates gravitationally. The most reliable measurements are those that resolve the smallest physical scales around the SMBHs.
  We consider here three metrics to compare the physical scales probed by kinematic tracers dominated by rotation: the radius of the innermost detected kinematic tracer $R_\mathrm{min}$ normalised by respectively the SMBH's Schwarzschild radius ($R_\mathrm{Schw}\equiv2GM_\mathrm{BH}/c^2$, where $G$ is the gravitational constant and $c$ the speed of light), sphere-of-influence (SOI) radius ($R_\mathrm{SOI}\equiv GM_\mathrm{BH}/\sigma_\mathrm{e}^2$, where $\sigma_\mathrm{e}$ is the stellar velocity dispersion within the galaxy's effective radius) and equality radius (the radius $R_\mathrm{eq}$ at which the SMBH mass equals the enclosed stellar mass, $M_\mathrm{BH}=M_*(R_\mathrm{eq})$, where $M_*(R)$ is the stellar mass enclosed within the radius $R$). All metrics lead to analogous simple relations between $R_\mathrm{min}$ and the highest circular velocity probed $V_\mathrm{c}$. Adopting these metrics to compare the SMBH mass measurements using molecular gas kinematics to those using megamaser kinematics, we demonstrate that the best molecular gas measurements resolve material that is physically closer to the SMBHs in terms of $R_\mathrm{Schw}$ but is slightly farther in terms of $R_\mathrm{SOI}$ and $R_\mathrm{eq}$. However, molecular gas observations of nearby galaxies using the most extended configurations of the Atacama Large Millimeter/sub-millimeter Array can resolve the SOI comparably well and thus enable SMBH mass measurements as precise as the best megamaser measurements.
\end{abstract}



\section{Introduction}
\label{sec:intro}

Supermassive black hole (SMBH) masses ($M_\mathrm{BH}$) correlate with several properties of their host galaxies, such as stellar velocity dispersion, bulge mass, and stellar mass \citep[e.g.][]{Magorrian_1998, Ferrarese_2000, Gebhardt_2000, Beifiori_2012, Kormendy_2013}. These correlations are sufficiently tight to suggest (potentially self-regulating) co-evolutionary processes, although their exact nature remains unclear \citep[e.g.][]{Kormendy_2013, Simmons_2017, Krajnovic_2018, Sahu_2019}.

The mass of a SMBH can be directly measured from observations of a kinematic tracer sufficiently close to it that its contribution to the total galactic gravitational potential can be disentangled from (i.e.\ is comparable to) the contributions of other constituents (e.g.\ stars, gas, dust and/or dark matter). This is simplest when observations resolve the spatial scales on which the SMBH dominates the potential \citep[e.g.][]{Boizelle_2019,North_2019,Ruffa_2023}. However, with a reliable model of the distributions of the other mass components, the SMBH's contribution can still be discerned (with commensurately larger uncertainties) by tracing velocities higher than those expected from models with no SMBH \citep[e.g.][]{Davis_2018, Smith_2019}. Stars and ionised gas are common kinematic tracers and have yielded most of the published SMBH mass measurements to date. Measurements with megamasers (hereafter "masers" for short) as tracers are often thought of as the "gold standard" \citep[e.g.][]{Herrnstein_2005, Kuo_2011, Gao_2017, Zhao_2018, Kuo_2020}, as very long baseline interferometry (VLBI) allows to resolve very small angular scales (and thus physical scales) close to the SMBHs.
However, suitable masers are rare and are detected almost exclusively in galaxies with Seyfert~2 active galactic nuclei (AGN), hosting SMBHs of relatively low masses ($10^6\lesssim M_\mathrm{BH}/\mathrm{M}_\odot\lesssim10^8$; e.g.\ \citealt{Braatz_1996, Greenhill_2003, Bosch_2016}).

More recently, advances in millimetre interferometry have enabled to resolve molecular gas on the spatial scales over which central SMBHs dominate the galactic potentials. Following an initial proof-of-concept measurement in NGC~4526 \citep{Davis_2013}, additional measurements have been performed by our millimetre-Wave Interferometric Survey of Dark Object Masses (WISDOM) in ten typical early-type galaxies (ETGs; \citealt{Onishi_2017, Davis_2017, Davis_2018, Smith_2019, North_2019, Smith_2021, Ruffa_2023, Zhang_2024_submitted, Dominiak_2024_submitted2}), a dwarf ETG \citep{Davis_2020}, and a peculiar luminous infrared galaxy (LIRG) with a Seyfert nucleus \citep{Lelli_2022}. Using similar techniques, other groups have presented molecular gas SMBH mass measurements of twelve additional ETGs \citep{Barth_2016, Boizelle_2019, Nagai_2019, Ruffa_2019, Boizelle_2021, Cohn_2021, Kabasares_2022, Nguyen_2022, Cohn_2023,  Dominiak_2024} and three late-type galaxies (LTGs), all barred spirals \citep{Onishi_2015, Nguyen_2020, Nguyen_2021}.
Although the molecular gas technique is well suited to a range of galaxies, it is more challenging to apply to LTGs because the shallower and occasionally non-axisymmetric potentials make non-circular gas motions more significant \citep[e.g.][]{Combes_2019}.

VLBI observations of masers for SMBH mass measurements resolve much smaller physical scales than molecular gas observations. However, the absolute scales do not necessarily offer a meaningful comparison, as the scale required to perform a SMBH mass measurement, the radius of the SMBH's sphere of influence (SOI), is proportional to the SMBH mass. \citet{Kormendy_2013} showed that, among all SMBH measurements published by then, that of NGC~4258 by \citet{Herrnstein_2005} using masers had the most resolution elements across the SOI and was thus the most precise extragalactic SMBH mass measurement. \citet{North_2019} however demonstrated that, considering instead the number of spatially-resolved Schwarzschild radii, their molecular gas observations of NGC~383 resolved spatial scales comparable to those of the maser measurements of the Megamaser Cosmology Project \citep[e.g.][]{Kuo_2011,Gao_2017}.

In this paper, we consider three separate metrics that quantify the physical scales probed by kinematic tracers in circular motions: the radius of the innermost detected dynamical tracer divided by respectively the SMBH's Schwarzschild radius $R_\mathrm{Schw}$, SOI radius $R_\mathrm{SOI}$ (estimated using the effective stellar velocity dispersion) and equality radius $R_\mathrm{eq}$ (the radius at which the SMBH mass equals the enclosed stellar mass). We then adopt the metrics to compare the resolved spatial scales of maser and molecular gas dynamical SMBH mass measurements. In Section~\ref{sec:theory}, we introduce the metrics by rewriting the circular velocities $V_\mathrm{c}(R)$ of the Keplerian circular velocity curves using different $M_\mathrm{BH}$-independent forms. In Section~\ref{sec:compare}, we estimate and collect the parameters of all existing SMBH mass measurements to date that use molecular gas and/or maser observations. We also compare their minimum resolved spatial scales using the metrics introduced and discuss the results and their implications. We conclude in Section~\ref{sec:conclude}.

\section{Keplerian circular velocity curves in different units}
\label{sec:theory}

For kinematic tracers dominated by rotation (so primarily molecular gas and masers rather than ionised gas or stars), the key factor for the precision of a SMBH mass measurement (assuming a sufficient signal-to-noise ratio has been achieved) is how well the observations resolve the innermost part of the rotation curve, where the motion is predominantly Keplerian and gravitationally-dominated by the SMBH \citep{Davis_2014}. In other words, a precise measurement requires observations with a spatial resolution sufficiently high to probe the kinematics very close to the SMBH and deep into its potential well. VLBI maser measurements are thus considered the gold standard of SMBH measurements, as they resolve spatial scales much smaller than those reached by other methods (see e.g. Table \ref{tab:all_meas}). However, as argued above, an absolute scale does not necessarily offer a meaningful metric, because the scale required to perform a SMBH mass measurement (the SOI) itself depends on $M_\mathrm{BH}$.
A better spatial resolution will only lead to a more precise $M_\mathrm{BH}$ measurement at a fixed mass. 
It is thus more meaningful to compare the spatial scales probed by different kinematic tracers using a set of units that eliminate the $M_\mathrm{BH}$ dependence. In this section, we thus rewrite the circular velocity $V_\mathrm{c}(R)$ of Keplerian rotation using three different pairs of such units and introduce three corresponding metrics to compare SMBH mass measurements that use different tracers.

\subsection{Natural units ($c$ and $R_\mathrm{Schw}$)}

The velocity of any kinematic tracer in a circular motion due to gravity is such that the gravitational acceleration equals the centrifugal acceleration:
\begin{equation}
  \frac{GM(R)}{R^2}=\frac{V_\mathrm{c}^2}{R},
  \label{eq:circular_motion}
\end{equation}
where $G$ is the gravitational constant, $M(R)$ the mass enclosed within a sphere of radius $R$ and $V_\mathrm{c}$ the circular velocity at $R$. 
The enclosed mass $M(R)$ includes contributions from all the mass components (SMBH, stars, gas, dust, dark matter, etc). However, for the rest of the discussion, we assume that $M(R)$ in the central region of each galaxy is dominated by the SMBH and the stars, as the gas fraction and the dark matter fraction are usually low in the nuclear regions of local galaxies \citep[e.g.][]{Cappellari_2013, Zhu_2024}. We also note that for a non-spherically symmetric mass distribution, mass outside of the radius $R$ also contributes to the kinematics. Those contributions however tend to cancel out, and the net effect is negligible compared to the high circular velocities in the nuclear region.
Multiplying both sides of equation~\eqref{eq:circular_motion} by the Schwarzschild radius
\begin{equation}
  R_\mathrm{Schw} \equiv \frac{2GM_\mathrm{BH}}{c^2}
\end{equation}
 \citep{Schwarzschild_1916}, where $c$ is the speed of light, yields
\begin{equation}
  \left(\frac{V_\mathrm{c}}{c}\right)^2=\frac{1}{2}\left(\frac{R}{R_\mathrm{Schw}}\right)^{-1}\frac{M(R)}{M_\mathrm{BH}}.
  \label{eq:Keplerian_natural_exact}
\end{equation}
For a tracer much closer to the SMBH than the SOI radius, the BH mass dominates over the stellar mass (i.e.\ $M(R)\approx M_\mathrm{BH}\,$), and
\begin{equation}
  \left(\frac{V_\mathrm{c}}{c}\right)^2 \approx \frac{1}{2}\left(\frac{R}{R_\mathrm{Schw}}\right)^{-1}\,\,\,.
  \label{eq:Keplerian_natural_appro}
\end{equation}
In this manner, we have rewritten the Keplerian relation between $V_\mathrm{c}$ and $R$ in natural units, eliminating its explicit dependence on the SMBH mass. This approach allows a fair comparison of observations of different kinematic tracers, in terms of the proximity of each tracer to the SMBH in units of the Schwarzschild radius, even when the SMBH masses are different by orders of magnitude. Observations that resolve circumnuclear disc material the fewest Schwarzschild radii away from the SMBH will thus detect the highest rotational velocities and reveal the physical processes of SMBH accretion and feedback closest to the galactic nucleus.

As a caveat, the Keplerian relation in equation~\eqref{eq:Keplerian_natural_appro} is only satisfied by kinematic tracers well within the SOI, so if an observation does not highly resolve the SOI, as is the case for many molecular gas observations (shown later in Section~\ref{subsec:RSOI}), the velocities of the kinematic tracer are expected to deviate from equation~\eqref{eq:Keplerian_natural_appro}. By contrast, most SMBH measurement observations using masers do highly resolve the SOI. The masers are thus expected to closely follow equation~\eqref{eq:Keplerian_natural_appro}. This will be discussed in detail in Section~\ref{subsec:RSOI}.

\subsection{Units of $\sigma_\textrm{e}$ and $R_\mathrm{SOI}$}

Equations~\eqref{eq:Keplerian_natural_exact} and \eqref{eq:Keplerian_natural_appro} express $V_\mathrm{c}$ and $R$ in natural units, providing a metric that quantifies how well an observation spatially resolves the material and motions in the circumnuclear disc.
However, this metric is not ideal for evaluating the reliability of SMBH mass measurements, as the natural units relate only to the SMBH and contain no information on how other components of the gravitational potential compare to it. For example, at a fixed radius in natural units, a kinematic tracer in a galaxy with an overly massive central bulge (compared to its SMBH) will be more affected by stars than one in a galaxy with an under-massive bulge. Everything else being equal, the latter case generally yields a more precise SMBH mass than the former. To evaluate a SMBH mass measurement in units that reflect both the SMBH and the stars, and thus the likely precision of the measurement, we normalise $R$ by the radius of the SMBH SOI
\begin{equation}
  R_\mathrm{SOI}\equiv\frac{GM_\mathrm{BH}}{\sigma_\mathrm{e}^{2}}
  \label{eq:SOI_def}
\end{equation}
\citep[e.g.][]{Wolfe_1970}, where $\sigma_\mathrm{e}$ is the stellar velocity dispersion measured within one effective (half-light) radius $R_\mathrm{e}$ of the galaxy, a proxy for the depth of the galaxy potential well and thus the stellar component at large spatial scales. The relation between $V_\mathrm{c}$ and $R$ then becomes
\begin{equation}
  \left(\frac{V_\mathrm{c}}{\sigma_\mathrm{e}}\right)^2=\left(\frac{R}{R_\mathrm{SOI}}\right)^{-1}\frac{M(R)}{M_\mathrm{BH}}.
  \label{eq:Keplerian_rsoi_exact}
\end{equation}
Again, for $R\ll R_\mathrm{SOI}$, $M(R)\approx M_\mathrm{BH}$, and
\begin{equation}
  \left(\frac{V_\mathrm{c}}{\sigma_\mathrm{e}}\right)^2 \approx \left(\frac{R}{R_\mathrm{SOI}}\right)^{-1}.
  \label{eq:Keplerian_rsoi_appro}
\end{equation}

 It is worth noting that when $R$ is normalised by $R_\mathrm{SOI}$, $V_\mathrm{c}$ is normalised by $\sigma_\mathrm{e}$. This is reasonable as the SOI is the region where the influence of the SMBH potential exceeds that of the stellar potential. So, for a tracer at $R_\mathrm{SOI}$, the SMBH's contribution to the circular velocity should approximately equal the contribution from the stars, which is of the order of $\sigma_\mathrm{e}$. Again, observations that only marginally resolve the SOI ($R\approx R_\mathrm{SOI}$) have $M(R)$ considerably larger than $M_\mathrm{BH}$, so the velocities are expected to deviate from equation~\eqref{eq:Keplerian_rsoi_appro}. 

\subsection{Units of $V_\mathrm{eq}$ and $R_\mathrm{eq}$}

Although $\sigma_\mathrm{e}$ is measurable for most nearby galaxies and provides a convenient way to estimate the size of the SOI, it is only a proxy for the dynamics of the stars and their underlying gravitational potential. The SOI definition in equation~\eqref{eq:SOI_def} is thus only an approximation to the more physically meaningful definition that the SOI is the region where the SMBH potential dominates over the stellar potential, or equivalently where the SMBH mass dominates over the stellar mass. Therefore, we consider a more formal and accurate definition of the SOI radius, the equality radius $R_\mathrm{eq}$, such that
\begin{equation}
    M_*(R_\mathrm{eq})=M_\mathrm{BH}\,\,\,,
\end{equation}
where $M_*(R)$ is the stellar mass enclosed within the radius $R$. We then define $V_\mathrm{eq}$ to be the circular velocity at $R_\mathrm{eq}$ due only to the SMBH:
\begin{equation}
    V_\mathrm{eq}\equiv\sqrt{\frac{GM_\mathrm{BH}}{R_\mathrm{eq}}}\,\,\,.
\end{equation}
In these units, the relation between $V_\mathrm{c}$ and $R$ becomes
\begin{equation}
    \left(\frac{V_\mathrm{c}}{V_\mathrm{eq}}\right)^2=\left(\frac{R}{R_\mathrm{eq}}\right)^{-1}\frac{M(R)}{M_\mathrm{BH}}\,\,\,.
    \label{eq:Keplerian_req_exact}
\end{equation}
Again, for $R\ll R_\mathrm{eq}$, $M(R)\approx M_\mathrm{BH}$, and
\begin{equation}
    \left(\frac{V_\mathrm{c}}{V_\mathrm{eq}}\right)^2 \approx \left(\frac{R}{R_\mathrm{eq}}\right)^{-1}\,\,\,.
    \label{eq:Keplerian_req_appro}
\end{equation}

Unlike the definition of $R_\mathrm{SOI}$ that approximates the comparison between the SMBH and the stellar potential with the comparison between $V_\mathrm{c}$ and $\sigma_\mathrm{e}$, the definition of $R_\mathrm{eq}$ comes from an explicit comparison between the SMBH mass and the stellar mass distribution. Therefore, $V_\mathrm{eq}$ and $R_\mathrm{eq}$ are better physically-motivated and more accurate units to compare the depths of the potentials probed by different observations. Having said that, detailed modelling of the stellar mass distribution near the SMBH SOI is not available (nor feasible) for every galaxy with a dynamically measured SMBH mass. This is particularly the case for maser galaxies, as their SOI are often smaller than the angular resolutions of even the best optical telescopes (e.g.\ {\sl Hubble Space Telescope}, {\sl HST}, and {\sl James Webb Space Telescope}; see Section~\ref{subsec:RSOI}). Hence, $V_\mathrm{eq}$ and $R_\mathrm{eq}$ are only available for a subset of the galaxies considered in the next section, and this will be discussed in detail.

\section{Comparing maser and molecular gas measurements}
\label{sec:compare}

\subsection{Methods}

In this section, we compile all molecular gas and maser SMBH mass measurements in the literature and compare them using the metrics introduced in Section~\ref{sec:theory}. Instead of re-deriving the entire rotation curve of each galaxy in the new units, we consider only the radius $R_\mathrm{min}$ and velocity $V_\mathrm{max}$ of the innermost kinematic tracer measurement in these units. This reveals the smallest spatial scale probed by each observation and indicates the quality of the mass measurement.

For maser galaxies, we thus adopt the (redshifted or blueshifted) maser spot that has the highest velocity relative to the galaxy's dynamical centre. The uncertainties of the maser position and velocity and those of the dynamical centre position and velocity are taken from the original publication and are added using the standard error propagation procedures for Gaussian random variables. 

By contrast, molecular gas observations yield continuous emission distributions in the radius-velocity planes, commonly shown in the literature as major-axis position-velocity diagrams (PVDs). For each galaxy, we thus visually identify the major-axis position (i.e.\ the radius) with the highest velocity (i.e.\ the velocity peak) in the central region of the PVD. If the observation resolves the SOI, this velocity traces the velocity rise due to the SMBH. If the SOI is unresolved and no velocity rise is detected, this velocity represents the innermost resolved velocity within which the observed velocity rapidly falls off to zero. In this case, a BH mass measurement is still possible if the velocity is substantially higher than that expected from models of the other mass components only.

In practice, we extract the corresponding radius and velocity using WebPlotDigitizer\footnote{\url{https://automeris.io/WebPlotDigitizer/index.html}} \citep{Rohatgi2022}. If the redshifted and blueshifted velocity peaks look almost symmetric, we measure both peaks and adopt the average $R_\mathrm{min}$ and $V_\mathrm{max}$ of the two measurements, with half of the differences as the uncertainties. If the resulting $V_\mathrm{max}$ uncertainty is smaller than the channel width of the data cube, we inflate it to the channel width. 

We note that the channel width is a conservative estimate of the typical uncertainty of velocity measurements from data cubes. 
If one of the peaks has a substantially lower velocity than the other or is indistinguishable from noise, we adopt the other (higher velocity) peak and estimate the position uncertainty from the spatial width of that peak and the velocity uncertainty as the channel width.
We also note that $R_\mathrm{min}$ is sometimes marginally smaller than the full-width at half-maximum (FWHM) of the synthesised beam of the observations, as the position of the central unresolved emission can be measured more precisely than the beam FWHM.
We show five examples of our measurements of $R_\mathrm{min}$ and $V_\mathrm{max}$ for molecular gas observations in Appendix~\ref{appendix_examples}. Three examples are from observations that resolve the SOI, and two are from observations that do not. The rest of the measurements are shown in the online supplementary material (PVDs reproduced with permission). Finally, we deproject the observed maximum rotation velocity $V_\mathrm{max}$ into the maximum circular velocity assuming perfect circular motion and an infinitely thin disc, $V_\mathrm{c}=V_\mathrm{max}/\sin{(i)}$, adopting the inclination and inclination uncertainty listed in each SMBH mass measurement paper.

While there may be non-circular motions or warps in some molecular gas and maser discs, in most cases the errors introduced in $V_\mathrm{c}$ will be small ($\lesssim30$~km~s$^{-1}$; see e.g.\ \citealt{Herrnstein_2005, Nguyen_2020}) compared to the large $V_\mathrm{c}$ in the nuclear regions. These errors are thus unimportant for the following discussion. More complex central non-circular motions (e.g.\ inflows and/or outflows) associated with bright emission could of course exist in some objects, and they could mimic Keplerian rises of rotation curves due to SMBHs. The galaxy Fairall~49 may be such an example, with faint blueshifted emission near the nucleus explainable by either a Keplerian velocity rise or non-circular motions (\citealt{Lelli_2022} argued for a molecular gas inflow). Our measurements of $V_\mathrm{max}$ and $R_\mathrm{min}$ for this object could thus be more significantly affected.

We thus acknowledge that our measurements of $R_\mathrm{min}$ and $V_\mathrm{max}$ of molecular gas observations are only rough estimates with considerable uncertainties. For example, the identification of the velocity peaks is performed visually and may be subject to human biases and uncertainties (e.g.\ confirmation bias and uncertainties marking the exact location of each peak by hand). Moreover, this method slightly but systematically overestimates $V_\mathrm{max}$, as it adopts the highest detected velocity, slightly boosted by the velocity dispersion of the gas, rather than the mean of the velocity distribution of the innermost kinematic tracer. We take these effects into account in the adopted uncertainties. For example, we repeat our measurements three times and compare the results to estimate human-related uncertainties. We also ensure that the uncertainty of $V_\mathrm{max}$ is at least as large as the gas velocity dispersion. The final adopted uncertainties are thus conservative and likely much larger than the systematic uncertainties caused by any of these effects. A more reliable method to measure the $R_\mathrm{min}$ and $V_\mathrm{max}$ of molecular gas observations could be to inspect the original data cube of each galaxy and identify the ``innermost detected kinematic tracer'' by applying a signal-to-noise ratio threshold. However, we have confirmed using the data cubes of the WISDOM project that these potentially more accurate measurements do not offer a substantial improvement over our more simplistic approach. 

\subsection{Results}

\subsubsection{Summary of measured and adopted quantities}

Table~\ref{tab:all_meas} lists all molecular gas and maser SMBH mass measurements in the literature, as well as our own measurements of $R_\mathrm{min}$ and $V_\mathrm{max}$ obtained from those. If a galaxy has multiple SMBH mass measurements from the same method, we consider only the most recent one. All uncertainties listed are $1\sigma$ uncertainties. Some inclinations have zero associated uncertainties as they were fixed during the kinematic modelling. If a maser paper does not state an inclination, we assume that the inclination was fixed at $90\degree$, as detection of maser emission requires an almost edge-on maser disc. We added the stellar dynamical mass measurement of Sgr~A* in the Milky Way (MW) as a reference point, as the motions of individual stars can be measured. The MW $R_\mathrm{min}$ and $V_\mathrm{max}$ are taken to be those of the star S2 from \citeauthor{GRAVITY_2018} (\citeyear{GRAVITY_2018}; hereafter \citetalias{GRAVITY_2018}). 

If uncertainties are provided, the distances listed in Table~\ref{tab:all_meas} are taken from the listed references. Otherwise, we adopt the distances listed in \citet{Saglia_2016} and re-scale the SMBH masses accordingly (as dynamical mass measurements scale linearly with the assumed distance). If a galaxy is not in \citet{Saglia_2016}, we adopt a distance from another source, retrieved from the HyperLeda\footnote{\url{http://leda.univ-lyon1.fr/}} database \citep{Makarov_2014}, and assume a $10\%$ uncertainty, consistent with the approach of \citet{Bosch_2016}. 

We also compile the stellar velocity dispersions within one effective radius $\sigma_\mathrm{e}$ of all the galaxies to compute their $R_\mathrm{SOI}$. For the MW, we adopt $\sigma_\mathrm{e}=105\pm20$~km~s$^{-1}$ \citep{Gultekin_2009}. For other galaxies, we prioritise the $\sigma_\mathrm{e}$ referenced by the SMBH mass measurement paper. If that paper does not list $\sigma_\mathrm{e}$, we search the following sources (in descending order of priority):
\begin{enumerate}
    \item Sources with well spatially-resolved stellar kinematics, such as the ATLAS$^\mathrm{3D}$ project \citep{Cappellari_2013} and the MASSIVE survey \citep{Veale_2017}. For MASSIVE galaxies, if uncertainties are not provided, we assume $10$~\kms.
    \item \citet{Bosch_2016}, who compiled the best $\sigma_\mathrm{e}$ measurements of all galaxies with a SMBH mass measurement before June 2016.
    \item Other literature sources \citep[e.g.][]{Greene_2010} with $\sigma_\mathrm{e}$ measurements.
    \item The "central stellar velocity dispersion" listed in HyperLeda \citep{Makarov_2014}, defined as the mean stellar velocity dispersion within a circular aperture of $0.595$~kpc radius.
    \item The "velocity dispersion" listed in the Sloan Digital Sky Survey (SDSS) data release 16 (DR16; \citealt{SDSS_2020}), measured within a circular aperture of $1.5$ arcsec radius.
\end{enumerate}

We correct the stellar velocity dispersions from HyperLeda and SDSS to $\sigma_\mathrm{e}$ using $\sigma_\mathrm{e}/\sigma=(R/R_\mathrm{e})^{0.08}$ (for spiral galaxies; \citealt{FB_2017}), where $R$ is the radius of the aperture and $\sigma$ the corresponding velocity dispersion, and $R_\mathrm{e}$ is computed from the stellar mass model adopted by the SMBH mass measurement paper. If the galaxy does not have a stellar mass model, we adopt the $r$-band Petrosian half-light radius from the SDSS DR16 \citep{SDSS_2020}. Because stellar velocity dispersion scales only slowly with aperture size ($\sigma_\mathrm{e}/\sigma=(R/R_\mathrm{e})^\alpha$, where $-0.06<\alpha< 0.08$; \citealt{FB_2017,Zhu_2024}), the differences between $\sigma$ and $\sigma_\mathrm{e}$ are unlikely to substantially affect our results. We are unaware of any $\sigma$ measurement of Fairall~49.

To measure $R_\mathrm{eq}$, we use the stellar mass model described in each SMBH mass measurement paper. For molecular gas measurements, this usually implies multiplying the best-fitting multi-Gaussian expansion (MGE; \citealt{EMB_1994, Cappellari_2002}) model of the stellar light distribution stated, constructed using the \texttt{mge\_fit\_sectors} procedure of \citet{Cappellari_2002}, with the corresponding best-fitting mass-to-light ratio ($M/L$). We then use the \texttt{mge\_radial\_mass} procedure\footnote{From \url{https://pypi.org/project/mgefit/}} in the \textsc{jampy} package\footnote{From \url{https://pypi.org/project/jampy/}} \citep{Cappellari_2008, Cappellari_2020} to convert the two-dimensional stellar mass profile into a one-dimensional radial profile. We interpolate this radial profile to identify the radius $R_\mathrm{eq}$ at which the enclosed stellar mass equals the best-fitting SMBH mass. Finally, we estimate the uncertainty of $R_\mathrm{eq}$ using Monte Carlo methods: we recompute $R_\mathrm{eq}$ using $10^4$ random realisations of $M_\mathrm{BH}$ and $M/L$ sampled from Gaussian distributions with means and standard deviations equal to the best-fitting $M_\mathrm{BH}$ and $M/L$ and their $1\sigma$ uncertainties, respectively. We take the standard deviation of the resultant distribution of $R_\mathrm{eq}$ as the uncertainty. The $R_\mathrm{eq}$ of the MW is calculated from the MW nuclear star cluster model of \citet{FK_2017}.

Our procedure to measure $R_\mathrm{eq}$ is reliable when the adopted stellar mass model comes from an overall dust-free optical or near-infrared image (usually from {\sl HST}) with a resolution comparable to or better than $R_\mathrm{eq}$. Otherwise, the result will depend strongly on the point-spread function (PSF) and/or the dust distribution model assumed. By checking the 
Mikulski Archive for Space Telescopes\footnote{\url{https://mast.stsci.edu/search/ui/\#/hst}}, 
we confirm that no image satisfies both criteria for any of the maser galaxies in this paper except NGC~4258 (whose model is detailed in \citealt{Drehmer_2015}). The main limitation is that maser galaxies typically have much smaller SOI ($R_\mathrm{eq}\sim0\farcs01$) than galaxies with molecular gas measurements (see Section~\ref{subsec:RSOI}). Additionally, the dense molecular gas required to trigger maser emission is often associated with prominent dust, so maser galaxies are typically dusty in their cores. By contrast, all of the SMBH mass measurement papers using molecular gas contain a reliable stellar mass model.

\begin{landscape}
\begin{table}
    \centering
    \caption{Molecular gas and maser SMBH mass measurements.}
    \label{tab:all_meas}
    \begin{tabular}{lcccccccccl} \hline
        \multirow{2}{2.5em}{Galaxy} & Distance & $R_\mathrm{min}$ & $R_\mathrm{min}$ & $V_\mathrm{max}$ & $\sigma_\mathrm{e}$ & Inclination & $\log{\left(M_\mathrm{BH}/\mathrm{M}_\odot\right)}$ & $R_\mathrm{SOI}$ & $R_\mathrm{eq}$ & \multirow{2}{4em}{Reference} \\
        & (Mpc) & (mas) & (pc) & (km $\rm s^{-1}$) & (km $\rm s^{-1}$) & (deg) & & (mas) & (mas) & \\ \hline
        \multicolumn{9}{l}{Molecular gas measurements} \\ \hline
        Fairall~49 & $\phantom{0}86.7 \pm \phantom{0}6.1$ & $\phantom{0}110\phantom{.00} \pm \phantom{0}30\phantom{.00}$ & $\phantom{0}46\phantom{.000} \pm 13\phantom{.000}$ & $\phantom{0}208 \pm 15$ & ... & $58\phantom{.0} \pm \phantom{0}3\phantom{.0}$ & $8.20 \pm 0.28$ & ... & $\phantom{0}280 \pm \phantom{0}90$ & \citet{Lelli_2022} \\
        NGC~315 & $\phantom{0}70\phantom{.1} \pm \phantom{0}7.0$ & $\phantom{00}70\phantom{.00} \pm \phantom{0}30\phantom{.00}$ & $\phantom{0}24\phantom{.000} \pm 10\phantom{.000}$ & $\phantom{0}532 \pm 10$ & $341 \pm \phantom{*}7$ & $74.2 \pm \phantom{0}0.1$ & $9.32 \pm 0.05$ & $230\phantom{.0} \pm \phantom{0}30\phantom{.0}$ & $\phantom{0}740 \pm \phantom{0}70$ & \citet{Boizelle_2021} \\
        NGC~383 & $\phantom{0}66.6 \pm \phantom{0}9.9$ & $\phantom{00}44\phantom{.00} \pm \phantom{00}7\phantom{.00}$ & $\phantom{0}14\phantom{.000} \pm \phantom{0}2\phantom{.000}$ & $\phantom{0}634 \pm 20$ & $239 \pm 16$ & $38\phantom{.0} \pm \phantom{0}1\phantom{.0}$ & $9.55 \pm 0.02$ & $840\phantom{.0} \pm 120\phantom{.0}$ & $\phantom{0}830 \pm \phantom{0}20$ & \citet{Zhang_2024_submitted} \\
        NGC~404 & $\phantom{00}3.1 \pm \phantom{0}0.4$ & $\phantom{00}75\phantom{.00} \pm \phantom{0}35\phantom{.00}$ & $\phantom{00}1.1\phantom{00} \pm \phantom{0}0.5\phantom{00}$ & $\phantom{00}61 \pm \phantom{0}2$ & $\phantom{*}40 \pm \phantom{*}3$ & $37\phantom{.0} \pm \phantom{0}1\phantom{.0}$ & $5.74 \pm 0.09$ & $100\phantom{.0} \pm \phantom{0}30\phantom{.0}$ & $\phantom{00}60 \pm \phantom{0}10$ & \citet{Davis_2020} \\ 
        NGC~524 & $\phantom{0}23.3 \pm \phantom{0}2.3$ & $\phantom{0}480\phantom{.00} \pm \phantom{0}20\phantom{.00}$ & $\phantom{0}54\phantom{.000} \pm \phantom{0}2\phantom{.000}$ & $\phantom{0}131 \pm 10 $ &$220 \pm 11$ & $20\phantom{.0} \pm \phantom{0}5\phantom{.0}$ & $8.60 \pm 0.22$ & $310\phantom{.0} \pm 160\phantom{.0}$ & $\phantom{0}280 \pm \phantom{0}70$ & \citet{Smith_2019} \\
        NGC~997 & $\phantom{0}90.4 \pm \phantom{0}9.0$ & $\phantom{0}190\phantom{.00} \pm \phantom{0}30\phantom{.00}$ & $\phantom{0}83\phantom{.000} \pm 13\phantom{.000}$ & $\phantom{0}162 \pm \phantom{0}5$ & $215 \pm 10$ & $35\phantom{.0} \pm \phantom{0}0\phantom{.0}$ & $8.99 \pm 0.25$ & $210\phantom{.0} \pm 120\phantom{.0}$ & $\phantom{0}250 \pm \phantom{0}60$ &\citet{Dominiak_2024} \\
        NGC~1097 & $\phantom{0}14.5 \pm \phantom{0}1.5$ & $1450\phantom{.00} \pm 250\phantom{.00}$ & $102\phantom{.000} \pm 18\phantom{.000}$ & $\phantom{0}210 \pm 10$ & $195 \pm \phantom{*}4$ & $46\phantom{.0} \pm \phantom{0}5\phantom{.0}$ & $8.15 \pm 0.10$ & $230\phantom{.0} \pm \phantom{0}50\phantom{.0}$ & $\phantom{0}530 \pm \phantom{0}40$ & \citet{Onishi_2015} \\
        NGC~1275 & $\phantom{0}71\phantom{.0} \pm \phantom{0}7.1$ & $\phantom{00}30\phantom{.00} \pm \phantom{0}10\phantom{.00}$ & $\phantom{0}10\phantom{.000} \pm \phantom{0}3\phantom{.000}$ & $\phantom{0}420 \pm 20$ & $245 \pm 28$ & $46\phantom{.0} \pm \phantom{0}9\phantom{.0}$ & $9.04 \pm 0.19$ & $230\phantom{.0} \pm 110\phantom{.0}$ & $\phantom{0}530 \pm 120$ & \citet{Nagai_2019} \\
        NGC~1332 & $\phantom{0}22.3 \pm \phantom{0}1.9$ & $\phantom{0}210\phantom{.00} \pm \phantom{0}50\phantom{.00}$ & $\phantom{0}31\phantom{.000} \pm \phantom{0}5\phantom{.000}$ & $\phantom{0}500 \pm 20$ & $331 \pm 15$ & $84\phantom{.0} \pm \phantom{0}1\phantom{.0}$ & $8.82 \pm 0.04$ & $240\phantom{.0} \pm \phantom{0}30\phantom{.0}$ & $\phantom{0}230 \pm \phantom{0}10$ & \citet{Barth_2016}\\
        NGC~1380 & $\phantom{0}17.1 \pm \phantom{0}1.7$ & $\phantom{0}150\phantom{.00} \pm \phantom{0}30\phantom{.00}$ & $\phantom{0}12\phantom{.000} \pm \phantom{0}2\phantom{.000}$ & $\phantom{0}270 \pm 10$ & $215 \pm \phantom{*}8$ & $76.9 \pm \phantom{0}0.1$ & $8.17 \pm 0.15$ &$170\phantom{.0} \pm \phantom{0}60\phantom{.0}$ & $\phantom{0}230 \pm \phantom{0}50$ & \citet{Kabasares_2022} \\
        NGC~1574 & $\phantom{0}19.3 \pm \phantom{0}1.9$ & $\phantom{00}60\phantom{.00} \pm \phantom{0}20\phantom{.00}$ & $\phantom{0}\phantom{0}5.6\phantom{00} \pm \phantom{0}1.9\phantom{00}$ & $\phantom{0}120 \pm 10$ & $216 \pm 16$ & $27\phantom{.0} \pm \phantom{0}2\phantom{.0}$ & $8.00 \pm 0.08$ & $100\phantom{.0} \pm \phantom{0}20\phantom{.0}$ & $\phantom{0}190 \pm \phantom{0}30$ & \citet{Ruffa_2023} \\
        NGC~1684 & $\phantom{0}62.8 \pm \phantom{0}2.3$ & $\phantom{0}420\phantom{.00} \pm \phantom{0}90\phantom{.00}$ & $128\phantom{.000} \pm 27\phantom{.000}$ & $\phantom{0}334 \pm 10$ & $262 \pm 10$ & $66.5 \pm \phantom{0}0.4$ & $9.16 \pm 0.08$ & $300\phantom{.0} \pm \phantom{0}60\phantom{.0}$ & $\phantom{0}350 \pm \phantom{0}40$ & \citet{Dominiak_2024} \\
        NGC~3258 & $\phantom{0}31.9 \pm \phantom{0}3.9$ & $\phantom{00}90\phantom{.00} \pm \phantom{0}30\phantom{.00}$ & $\phantom{0}14\phantom{.000} \pm \phantom{0}5\phantom{.000}$ & $\phantom{0}486 \pm 10$ & $260 \pm 10$ & $49.0 \pm \phantom{0}0.1$ & $9.35 \pm 0.04$ & $920\phantom{.0} \pm 100\phantom{.0}$ & $\phantom{0}900 \pm \phantom{0}40$ & \citet{Boizelle_2019} \\
        NGC~3504 & $\phantom{0}32.4 \pm \phantom{0}2.1$ & $\phantom{00}75\phantom{.00} \pm \phantom{0}35\phantom{.00}$ & $\phantom{0}12\phantom{.000} \pm \phantom{0}5\phantom{.000}$ & $\phantom{0}148 \pm 10$ & $119 \pm 10$ & $53\phantom{.0} \pm \phantom{0}1\phantom{.0}$ & $7.19 \pm 0.05$ & $\phantom{0}30\phantom{.0} \pm \phantom{00}6\phantom{.0}$ & $\phantom{00}58 \pm \phantom{00}3$ & \citet{Nguyen_2020} \\
        NGC~3557 & $\phantom{0}43.3 \pm \phantom{0}4.3$ & $\phantom{0}140\phantom{.00} \pm \phantom{0}10\phantom{.00}$ & $\phantom{0}29\phantom{.000} \pm \phantom{0}2\phantom{.000}$ & $\phantom{0}225 \pm 22$ & $282 \pm 16$ & $56\phantom{.0} \pm \phantom{0}1\phantom{.0}$ & $8.85 \pm 0.02$ & $180\phantom{.0} \pm \phantom{0}20\phantom{.0}$ & $\phantom{0}620 \pm \phantom{0}10$ & \citet{Ruffa_2019} \\
        NGC~3593 & $\phantom{00}7\phantom{.0} \pm \phantom{0}2\phantom{.0}$ & $\phantom{0}290\phantom{.00} \pm \phantom{0}50\phantom{.00}$ & $\phantom{00}9.8\phantom{00} \pm \phantom{0}1.7\phantom{00}$ & $\phantom{0}124 \pm 10$ & \phantom{*}$55 \pm \phantom{*}7$ & $75.0 \pm \phantom{0}0.1$ & $6.38 \pm 0.08$ & $100\phantom{.0} \pm \phantom{0}30\phantom{.0}$ & $\phantom{00}90 \pm \phantom{0}10$ & \citet{Nguyen_2022} \\
        NGC~3665 & $\phantom{0}34.7 \pm \phantom{0}6.8$ & $\phantom{0}250\phantom{.00} \pm \phantom{0}50\phantom{.00}$ & $\phantom{0}42\phantom{.000} \pm \phantom{0}8\phantom{.000}$ & $\phantom{0}220 \pm 10$ & $219 \pm 10$ & $69.9 \pm \phantom{0}0.2$ & $8.76 \pm 0.03$ & $310\phantom{.0} \pm \phantom{0}40\phantom{.0}$ & $\phantom{0}540 \pm \phantom{0}20$ & \citet{Onishi_2017} \\
        NGC~4261 & $\phantom{0}31.1 \pm \phantom{0}3.1$ & $\phantom{00}65\phantom{.00} \pm \phantom{0}40\phantom{.00}$ & $\phantom{00}9.8\phantom{00} \pm \phantom{0}6.0\phantom{00}$ & $\phantom{0}582 \pm 15$ & $263 \pm 12$ & $60.8 \pm \phantom{0}0\phantom{.0}$ & $9.21 \pm 0.01$ & $670\phantom{.0} \pm \phantom{0}60\phantom{.0}$ & $1140 \pm \phantom{0}10$ & \citet{Ruffa_2023} \\
        NGC~4429 & $\phantom{0}16.5 \pm \phantom{0}1.6$ & $\phantom{0}550\phantom{.00} \pm 100\phantom{.00}$ & $\phantom{0}44\phantom{.000} \pm \phantom{0}8\phantom{.000}$ & $\phantom{0}170 \pm 10$ & $178 \pm \phantom{*}8$ & $66.8 \pm \phantom{0}0.1$ & $8.18 \pm 0.09$ & $260\phantom{.0} \pm \phantom{0}60\phantom{.0}$ & $\phantom{0}270 \pm \phantom{0}20$ & \citet{Davis_2018} \\
        NGC~4526 & $\phantom{0}16.4 \pm \phantom{0}1.8$ & $\phantom{0}350\phantom{.00} \pm \phantom{0}30\phantom{.00}$ & $\phantom{0}28\phantom{.000} \pm \phantom{0}2\phantom{.000}$ & $\phantom{0}288 \pm 10$ & $209 \pm 10$ & $79\phantom{.0} \pm \phantom{0}0\phantom{.0}$ & $8.65 \pm 0.14$ & $550\phantom{.0} \pm 180\phantom{.0}$ & $\phantom{0}540 \pm 100$ & \citet{Davis_2013} \\ 
        NGC~4697 & $\phantom{0}11.4 \pm \phantom{0}1.1$ & $\phantom{0}210\phantom{.00} \pm \phantom{0}30\phantom{.00}$ & $\phantom{0}12\phantom{.000} \pm \phantom{0}2\phantom{.000}$ & $\phantom{0}230 \pm 10$ & $169 \pm \phantom{*}8$ & $76\phantom{.0} \pm \phantom{0}1\phantom{.0}$ & $8.10 \pm 0.02$ & $340\phantom{.0} \pm \phantom{0}30\phantom{.0}$ & $\phantom{0}410 \pm \phantom{0}10$ & \citet{Davis_2017} \\
        NGC~4751 & $\phantom{0}26.9 \pm \phantom{0}2.9$ & $\phantom{0}210\phantom{.00} \pm \phantom{0}70\phantom{.00}$ & $\phantom{0}27\phantom{.000} \pm \phantom{0}9\phantom{.000}$ & $\phantom{0}687 \pm 30$ & $355 \pm 14$ & $78.6 \pm \phantom{0}0.2$ & $9.45 \pm 0.04$ & $740\phantom{.0} \pm \phantom{0}90\phantom{.0}$ & $\phantom{0}460 \pm \phantom{0}40$ & \citet{Dominiak_2024_submitted2} \\
        NGC~4786 & $\phantom{0}62.1 \pm \phantom{0}9.3$ & $\phantom{0}170\phantom{.00} \pm \phantom{0}30\phantom{.00}$ & $\phantom{0}51\phantom{.000} \pm \phantom{0}9\phantom{.000}$ & $\phantom{0}263 \pm 20$ & $264 \pm 10$ & $69.3 \pm \phantom{0}0.7$ & $8.70 \pm 0.14$ & $100\phantom{.0} \pm \phantom{0}30\phantom{.0}$ & $\phantom{0}220 \pm \phantom{0}30$ & \citet{Kabasares_2024_arxiv} \\
        NGC~5193 & $\phantom{0}45.7 \pm \phantom{0}3.2$ & $\phantom{0}170\phantom{.00} \pm \phantom{0}20\phantom{.00}$ & $\phantom{0}38\phantom{.000} \pm \phantom{0}4\phantom{.000}$ & $\phantom{0}234 \pm 10$ & $194 \pm \phantom{0}7$ & $60.7 \pm \phantom{0}0.1$ & $8.15 \pm 0.18$ & $\phantom{0}70\phantom{.0} \pm \phantom{0}30\phantom{.0}$ & $\phantom{00}90 \pm \phantom{0}20$ & \citet{Kabasares_2024_arxiv} \\
        NGC~6861 & $\phantom{0}27.3 \pm \phantom{0}4.5$ & $1390\phantom{.00} \pm 250\phantom{.00}$ & $184\phantom{.000} \pm 33\phantom{.000}$ & $\phantom{0}468 \pm 20$ & $389 \pm \phantom{*}3$ & $72.7 \pm \phantom{0}0.1$ & $9.30 \pm 0.20$ & $430\phantom{.0} \pm 200\phantom{.0}$ & $\phantom{0}500 \pm 140$ & \citet{Kabasares_2022} \\
        NGC~7052 & $\phantom{0}69.3 \pm \phantom{0}5.0$ & $\phantom{0}220\phantom{.00} \pm \phantom{0}20\phantom{.00}$ & $\phantom{0}74\phantom{.000} \pm \phantom{0}7\phantom{.000}$ & $\phantom{0}382 \pm 30$ & $266 \pm 13$ & $75\phantom{.0} \pm \phantom{0}1\phantom{.0}$ & $9.40 \pm 0.02$ & $450\phantom{.0} \pm \phantom{0}50\phantom{.0}$ & $\phantom{0}580 \pm \phantom{0}10$ & \citet{Smith_2021} \\
        NGC~7469 & $\phantom{0}68.4 \pm 18.8$ & $\phantom{0}160\phantom{.00} \pm \phantom{0}40\phantom{.00}$ & $\phantom{0}46\phantom{.000} \pm 13\phantom{.000}$ & $\phantom{0}182 \pm 10$ & $152 \pm 16$ & $53\phantom{.0} \pm \phantom{0}1\phantom{.0}$ & $7.20 \pm 0.97$ & $\phantom{00}9\phantom{.0} \pm \phantom{00}8\phantom{.0}$ & $\phantom{00}80 \pm \phantom{0}70$ & \citet{Nguyen_2021} \\
        PGC~11179 & $\phantom{0}89\phantom{.0} \pm \phantom{0}8.9$ & $\phantom{0}210\phantom{.00} \pm \phantom{0}10\phantom{.00}$ & $\phantom{0}91\phantom{.000} \pm \phantom{0}4\phantom{.000}$ & $\phantom{0}391 \pm 20$ & $266 \pm \phantom{0}9$ & $60.0 \pm \phantom{0}0.1$ & $9.28 \pm 0.09$ & $270\phantom{.0} \pm \phantom{0}60\phantom{.0}$ & $\phantom{0}140 \pm \phantom{0}20$ &  \citet{Cohn_2023} \\
        UGC~2698 & $\phantom{0}91\phantom{.0} \pm \phantom{0}9.1$ & $\phantom{0}120\phantom{.00} \pm \phantom{0}30\phantom{.00}$ & $\phantom{0}53\phantom{.000} \pm 13\phantom{.000}$ & $\phantom{0}481 \pm 20$ & $304 \pm \phantom{*}6$ & $68\phantom{.0} \pm \phantom{0}1\phantom{.0}$ & $9.39 \pm 0.14$ & $260\phantom{.0} \pm \phantom{0}80\phantom{.0}$ & $\phantom{0}170 \pm \phantom{0}30$ & \citet{Cohn_2021} \\ \hline
        \multicolumn{9}{l}{Maser measurements} \\ \hline
        CGCG~074-064 & $\phantom{0}87.6 \pm \phantom{0}7.6$ & $\phantom{000}0.28 \pm \phantom{00}0.01$ & $\phantom{00}0.119 \pm \phantom{0}0.004$ & $\phantom{0}902 \pm \phantom{0}3$ & $113 \pm \phantom{0}3$ & $91\phantom{.0} \pm \phantom{0}1\phantom{.0}$ & $7.38 \pm 0.04$ & $\phantom{0}24\phantom{.0} \pm \phantom{00}3\phantom{.0}$ & ... & \citet{Pesce_2020} \\
        Circinus & $\phantom{00}2.8 \pm \phantom{0}0.5$ & $\phantom{000}5.3\phantom{0} \pm \phantom{00}0.1\phantom{0}$ & $\phantom{00}0.072 \pm \phantom{0}0.001$ & $\phantom{0}260 \pm \phantom{0}2$ & $158 \pm 18$ & $90\phantom{.0} \pm \phantom{0}0\phantom{.0}$ & $6.06 \pm 0.08$ & $\phantom{0}15\phantom{.0} \pm \phantom{00}4\phantom{.0}$ & ... & \citet{Greenhill_2003} \\
        ESO~558–G009 & $107.6 \pm \phantom{0}5.9$ & $\phantom{000}0.44  \pm \phantom{00}0.09$ & $\phantom{00}0.23\phantom{0} \pm \phantom{0}0.05\phantom{0}$ & $\phantom{0}575 \pm 14$ & $170 \pm 20$ & $98\phantom{.0} \pm \phantom{0}1\phantom{.0}$ & $7.23 \pm 0.03$ & $\phantom{00}5\phantom{.0} \pm \phantom{00}1\phantom{.0}$ & ... & \citet{Gao_2017} \\ 
        IC~1481 & $\phantom{0}78.7 \pm \phantom{0}5.5$ & $\phantom{000}7.3\phantom{0} \pm \phantom{00}2.7\phantom{0}$ & $\phantom{00}2.8\phantom{00} \pm \phantom{0}1.0\phantom{00}$ & $\phantom{0}170 \pm \phantom{0}6$ & $\phantom{*}95 \pm 26$ & $90\phantom{.0} \pm \phantom{0}0\phantom{.0}$ & $7.11 \pm 0.11$ & $\phantom{0}16\phantom{.0} \pm \phantom{0}10\phantom{.0}$ & ... & \citet{Mamyoda_2009} \\
        IC~2560 & $\phantom{0}31\phantom{.0} \pm 13\phantom{.0}$ & $\phantom{000}1.14 \pm \phantom{00}0.17$ & $\phantom{00}0.17\phantom{0} \pm \phantom{0}0.03\phantom{0}$ & $\phantom{0}325 \pm 20$ & $141 \pm 10$ & $90\phantom{.0} \pm \phantom{0}0\phantom{.0}$ & $6.62 \pm 0.06$ & $\phantom{00}6\phantom{.0} \pm \phantom{00}1\phantom{.0}$ & ... & \citet{Yamauchi_2012} \\
        IRAS~08452-0011 & $213\phantom{.0} \pm 15\phantom{.0}$ & $\phantom{000}0.30 \pm \phantom{00}0.13$ & $\phantom{00}0.31\phantom{0} \pm \phantom{0}0.13\phantom{0}$ & $\phantom{0}832 \pm \phantom{0}4$ & $137 \pm 11$ & $85.2 \pm \phantom{0}0.3$ & $7.52 \pm 0.03$ & $\phantom{00}7\phantom{.0} \pm \phantom{00}1\phantom{.0}$ & ... & \citet{Kuo_2020} \\
        J0437+2456 & $\phantom{0}65.3 \pm \phantom{0}3.6$ & $\phantom{000}0.13 \pm \phantom{00}0.08$ & $\phantom{00}0.041 \pm \phantom{0}0.025$ & $\phantom{0}442 \pm 11$ & $110 \pm 13$ & $81\phantom{.0} \pm \phantom{0}1\phantom{.0}$ & $6.46 \pm 0.05$ & $\phantom{00}3\phantom{.0} \pm \phantom{00}1\phantom{.0}$ & ... & \citet{Gao_2017} \\
        J1346+5228 & $121\phantom{.0} \pm 12\phantom{.0}$ & $\phantom{000}0.22 \pm \phantom{00}0.04$ & $\phantom{00}0.13\phantom{0} \pm \phantom{0}0.02\phantom{0}$ & $\phantom{0}788 \pm \phantom{0}8$ & $139 \pm \phantom{*}4$ & $90\phantom{.0} \pm \phantom{0}0\phantom{.0}$ & $7.24 \pm 0.06$ & $\phantom{00}7\phantom{.0} \pm \phantom{00}1\phantom{.0}$ & ... & \citet{Zhao_2018} \\
        Mrk~1 & $\phantom{0}61.5 \pm \phantom{0}2.3$ & $\phantom{000}1.48 \pm \phantom{00}0.21$ & $\phantom{00}0.44\phantom{0} \pm \phantom{0}0.06\phantom{0}$ & $\phantom{0}203 \pm \phantom{0}4$ & $\phantom{*}86 \pm \phantom{*}4$  & $90\phantom{.0} \pm \phantom{0}2\phantom{.0}$ & $6.51 \pm 0.07$ & $\phantom{00}6\phantom{.0} \pm \phantom{00}1\phantom{.0}$ & ... & \citet{Kuo_2020} \\
        Mrk ~1029 & $120.8 \pm \phantom{0}6.6$ & $\phantom{000}0.39 \pm \phantom{00}0.05$ & $\phantom{00}0.23\phantom{0} \pm \phantom{0}0.03\phantom{0}$ & $\phantom{0}163 \pm 26$ & $132 \pm 15$ & $79\phantom{.0} \pm \phantom{0}2\phantom{.0}$ & $6.28 \pm 0.12 $ & $\phantom{00}0.8 \pm \phantom{00}0.3$ & ... & \citet{Gao_2017} \\
        NGC~1068 & $\phantom{0}15.9 \pm \phantom{0}9.4$ & $\phantom{000}8.0\phantom{0} \pm \phantom{00}0.2\phantom{0}$ & $\phantom{00}0.62\phantom{0} \pm \phantom{0}0.02\phantom{0}$ & $\phantom{0}330 \pm 10$ & $151 \pm \phantom{*}7$ & $90\phantom{.0} \pm \phantom{0}0\phantom{.0}$ & $6.95 \pm 0.02$ & $\phantom{0}22\phantom{.0} \pm \phantom{00}2\phantom{.0}$ & ... & \citet{Lodato_2003} \\
        NGC~1194 & $\phantom{0}58.0 \pm \phantom{0}6.3$ & $\phantom{000}2.17 \pm \phantom{00}0.30$ & $\phantom{00}0.61\phantom{0} \pm \phantom{0}0.08\phantom{0}$ & $\phantom{0}685 \pm 15$ & $148 \pm 24$ & $85\phantom{.0} \pm \phantom{0}0\phantom{.0}$ & $7.85 \pm 0.02$ & $\phantom{0}50\phantom{.0} \pm \phantom{0}20\phantom{.0}$ & ... & \citet{Kuo_2011} \\
        NGC~1320 & $\phantom{0}34.2 \pm \phantom{0}1.9$ & $\phantom{000}0.38 \pm \phantom{00}0.07$ & $\phantom{00}0.063 \pm \phantom{0}0.012$ & $\phantom{0}402 \pm 11$ & $141 \pm 16$ & $90\phantom{.0} \pm \phantom{0}0\phantom{.0}$ & $6.74 \pm 0.21$ & $\phantom{00}7\phantom{.0} \pm \phantom{00}4\phantom{.0}$ & ... & \citet{Gao_2017} \\
        NGC~2273 & $\phantom{0}29.5 \pm \phantom{0}1.9$ & $\phantom{000}0.28 \pm \phantom{00}0.05$ & $\phantom{00}0.040 \pm \phantom{0}0.007$ & $\phantom{0}958 \pm 15$ & $145 \pm 17$ & $84\phantom{.0} \pm \phantom{0}0\phantom{.0}$ & $6.93 \pm 0.02$ & $\phantom{0}12\phantom{.0} \pm \phantom{00}3\phantom{.0}$ & ... & \citet{Kuo_2011} \\
    \end{tabular}
\end{table}
\end{landscape}

\begin{landscape}
\begin{table}
\centering
\contcaption{}
 \begin{tabular}{lcccccccccl} \hline
        \multirow{2}{2.5em}{Galaxy} & Distance & $R_\mathrm{min}$ & $R_\mathrm{min}$ & $V_\mathrm{max}$ & $\sigma_\mathrm{e}$ & Inclination & $\log{\left(M_\mathrm{BH}/\mathrm{M}_\odot\right)}$ & $R_\mathrm{SOI}$ & $R_\mathrm{eq}$ & \multirow{2}{4em}{Reference} \\
        & (Mpc) & (mas) & (pc) & (km $\rm s^{-1}$) & (km $\rm s^{-1}$) & (deg) & & (mas) & (mas) & \\ \hline
        NGC~2960 & $\phantom{0}67.1 \pm \phantom{0}7.1$ & $\phantom{000}0.42 \pm \phantom{00}0.06$ & $\phantom{00}0.14\phantom{0} \pm \phantom{0}0.02\phantom{0}$ & $\phantom{0}581 \pm 15$ & $151 \pm \phantom{*}7$ & $89\phantom{.0} \pm \phantom{0}0\phantom{.0}$ & $7.03 \pm 0.02$ & $\phantom{00}6.2 \pm \phantom{00}0.6$ & ... & \citet{Kuo_2011} \\
        NGC~3079 & $\phantom{0}15.9 \pm \phantom{0}1.2$ & $\phantom{000}4.8\phantom{0} \pm \phantom{00}0.4\phantom{0}$ & $\phantom{00}0.37\phantom{0} \pm \phantom{0}0.03\phantom{0}$ & $\phantom{0}145 \pm 25$ & $145 \pm \phantom{*}7$ & $84\phantom{.0} \pm \phantom{0}0\phantom{.0}$ & $6.36 \pm 0.09$ & $\phantom{00}6\phantom{.0} \pm \phantom{00}1\phantom{.0}$ & ... & \citet{Yamauchi_2004} \\
        NGC~3393 & $\phantom{0}49.2 \pm \phantom{0}8.2$ & $\phantom{000}1.7\phantom{0} \pm \phantom{00}0.4\phantom{0}$ & $\phantom{00}0.41\phantom{0} \pm \phantom{0}0.10\phantom{0}$ & $\phantom{0}575 \pm \phantom{0}5$ & $148 \pm 10$ & $90\phantom{.0} \pm \phantom{0}0\phantom{.0}$ & $7.20 \pm 0.32$ & $\phantom{0}13\phantom{.0} \pm \phantom{0}10\phantom{.0}$ & ... & \citet{Kondratko_2008} \\
        NGC~4258\phantom{S0-000} & $\phantom{00}7.3 \pm \phantom{0}0.5$ & $\phantom{000}4.5\phantom{0} \pm \phantom{00}0.2\phantom{0}$ & $\phantom{00}0.159 \pm \phantom{0}0.007$ & $1001 \pm \phantom{0}7$ & $115 \pm 11$ & $81\phantom{.0} \pm \phantom{0}0\phantom{.0}$ & $7.58 \pm 0.01$ & $350\phantom{.0} \pm \phantom{0}70\phantom{.0}$ & $\phantom{0}244 \pm \phantom{00}5$ & \citet{Herrnstein_2005}\phantom{x} \\
         NGC~4388 & $\phantom{0}16.5 \pm \phantom{0}1.6$ & $\phantom{000}2.5\phantom{0} \pm \phantom{00}0.2\phantom{0}$ & $\phantom{00}0.20\phantom{0} \pm \phantom{0}0.02\phantom{0}$ & $\phantom{0}405 \pm \phantom{0}3$ & $107 \pm \phantom{*}7$ & $90\phantom{.0} \pm \phantom{0}0\phantom{.0}$ & $6.86 \pm 0.01$ & $\phantom{0}34\phantom{.0} \pm \phantom{00}5\phantom{.0}$ & ... & \citet{Kuo_2011} \\
        NGC~4945 & $\phantom{00}3.7 \pm \phantom{0}0.4$ & $\phantom{000}9.2\phantom{0} \pm \phantom{00}3.8\phantom{0}$ & $\phantom{00}0.17\phantom{0} \pm \phantom{0}0.07\phantom{0}$ & $\phantom{0}149 \pm \phantom{0}4$ & $135 \pm \phantom{*}6$ & $90\phantom{.0} \pm \phantom{0}0\phantom{.0}$ & $6.14 \pm 0.18$ & $\phantom{0}18\phantom{.0} \pm \phantom{00}8\phantom{.0}$ & ... & \citet{Greenhill_1997} \\
        NGC~5495 & $\phantom{0}95.7 \pm \phantom{0}5.3$ & $\phantom{000}0.37 \pm \phantom{00}0.15$ & $\phantom{00}0.17\phantom{0} \pm \phantom{0}0.07\phantom{0}$ & $\phantom{0}533 \pm 67$ & $166 \pm 19$ & $95\phantom{.0} \pm \phantom{0}1\phantom{.0}$ & $7.04 \pm 0.08$ & $\phantom{00}4\phantom{.0} \pm \phantom{00}1\phantom{.0}$ & ... & \citet{Gao_2017} \\ 
        NGC~5765B & $112.2 \pm \phantom{0}5.3$ & $\phantom{000}0.56 \pm \phantom{00}0.03$ & $\phantom{00}0.30\phantom{0} \pm \phantom{0}0.02\phantom{0}$ & $\phantom{0}760 \pm \phantom{0}3$ & $158 \pm 18$ & $72.4 \pm \phantom{0}0.5$ & $7.62 \pm 0.02$ & $\phantom{0}13\phantom{.0} \pm \phantom{00}3\phantom{.0}$ & ... & \citet{Pesce_2020b} \\ 
        NGC~6264 & $132.1 \pm 19.3$ & $\phantom{000}0.39 \pm \phantom{00}0.01$ & $\phantom{00}0.250 \pm \phantom{0}0.006$ & $\phantom{0}729 \pm \phantom{0}1$ & $159 \pm 15$ & $91\phantom{.0} \pm \phantom{0}2\phantom{.0}$ & $7.44 \pm 0.06$ & $\phantom{00}7\phantom{.0} \pm \phantom{00}2\phantom{.0}$ & ... & \citet{Pesce_2020b} \\ 
        NGC~6323 & $109.4 \pm 28.8$ & $\phantom{000}0.27 \pm \phantom{00}0.02$ & $\phantom{00}0.14\phantom{0} \pm \phantom{0}0.01\phantom{0}$ & $\phantom{0}540 \pm \phantom{0}2$ & $158 \pm 26$ & $91.5 \pm \phantom{0}0.3$ & $7.01 \pm 0.11$ & $\phantom{00}3\phantom{.0} \pm \phantom{00}1\phantom{.0}$ & ... & \citet{Pesce_2020b} \\
        UGC~3789 & $\phantom{0}51.5 \pm \phantom{0}4.3$ & $\phantom{000}0.35 \pm \phantom{00}0.01$ & $\phantom{00}0.087 \pm \phantom{0}0.002$ & $\phantom{0}847 \pm \phantom{0}1$ & $107 \pm 12$ & $84.9 \pm \phantom{0}0.6$ & $7.08 \pm 0.03$ & $\phantom{0}18\phantom{.0} \pm \phantom{00}4\phantom{.0}$ & ... & \citet{Pesce_2020b} \\
        UGC~6093 & $152\phantom{.0} \pm 15\phantom{.0}$ & $\phantom{000}0.29 \pm \phantom{00}0.05$ & $\phantom{00}0.21\phantom{0} \pm \phantom{0}0.04\phantom{0}$ & $\phantom{0}816 \pm 13$ & $155 \pm 18$ & $94\phantom{.0} \pm \phantom{0}4\phantom{.0}$ & $7.41 \pm 0.02$ & $\phantom{00}6\phantom{.0} \pm \phantom{00}2\phantom{.0}$ & ... & \citet{Zhao_2018} \\ \hline
        Sgr~A$^*$ (S2) & $0.008$ & $1400\,R_\mathrm{Schw}$ & $5.5\times10^{-4}$ & $7650$ & $105 \pm 20$ & ... & $6.61 \pm 0.01$ & $1.6$~pc & $3.1$~pc &  \citetalias{GRAVITY_2018} \\ \hline
\end{tabular}
\end{table}
\end{landscape}

\clearpage

\subsubsection{Sample galaxies in the $M_\mathrm{BH}$ -- $\sigma_\mathrm{e}$ diagram}

\begin{figure}
    \centering
    \includegraphics[width=\linewidth]{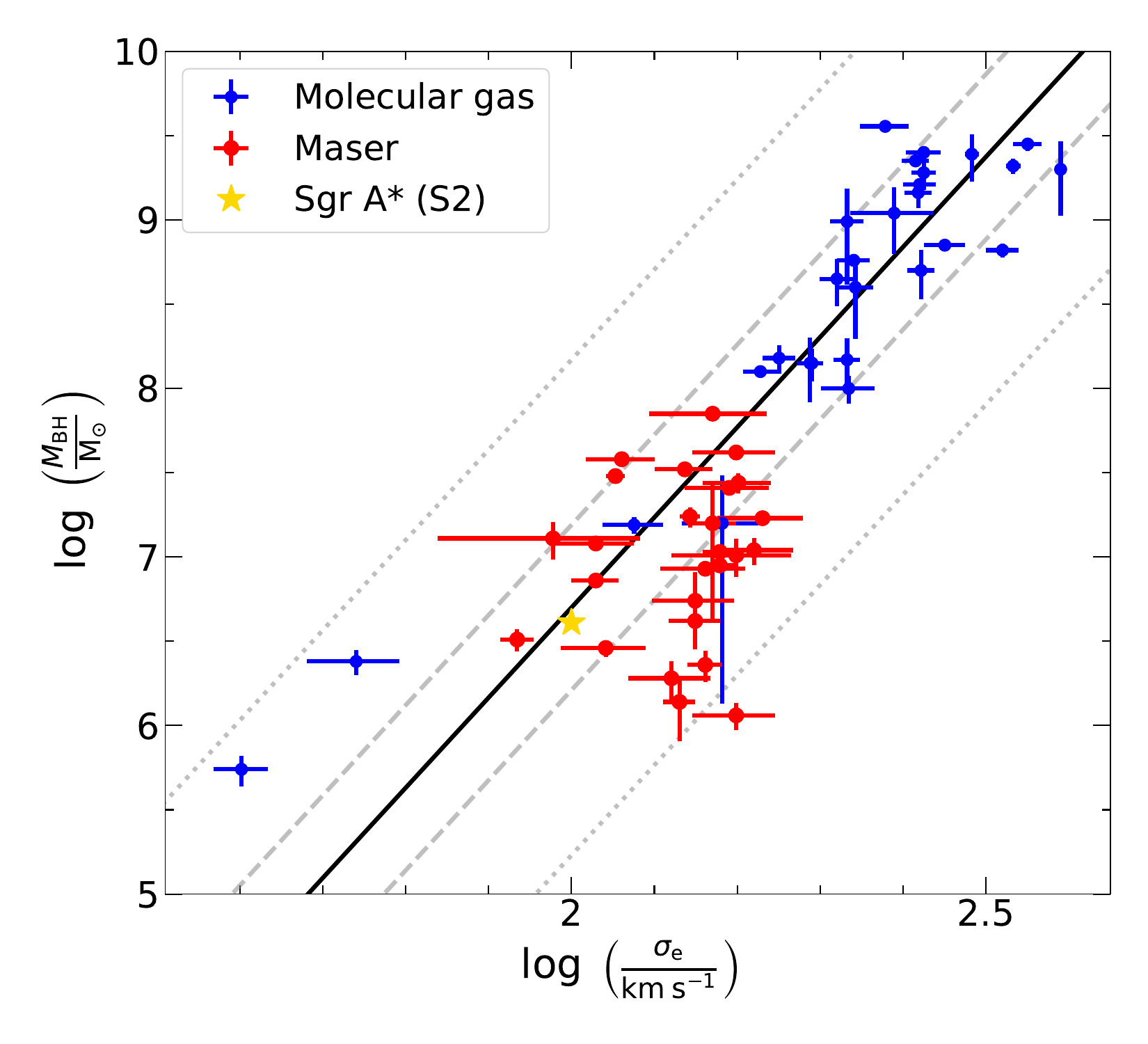}
    \caption{Correlation of the SMBH masses and the effective stellar velocity dispersions of all galaxies with a SMBH mass measurement using maser or molecular gas kinematics (and the MW). The solid black line shows the local $M_\mathrm{BH}$ -- $\sigma_\mathrm{e}$ relation of \citet{Bosch_2016}, while the dashed and dotted grey lines show the $1\sigma$ and $3\sigma$ observed scatter of the relation, respectively. Maser observations (filled red circles) probe a narrow SMBH mass range ($M_\mathrm{BH}\sim10^7$~M$_{\odot}$), much narrower than that of molecular gas observations (filled blue circles).
    }
    \label{fig:M-sigma}
\end{figure}

Figure~\ref{fig:M-sigma} shows the correlation of the SMBH masses and the effective stellar velocity dispersions of all galaxies with a SMBH mass measurement using maser or molecular gas kinematics (and the MW), compared to the local $M_\mathrm{BH}$ -- $\sigma_\mathrm{e}$ relation of \citet{Bosch_2016}. Maser observations span a SMBH mass range ($M_\mathrm{BH}\sim10^7$~M$_{\odot}$) much narrower than that of molecular gas observations. The main reason is that the required maser emission originates from a specific type of nuclear activity most frequently present in Seyfert~2 AGN (see \citealt{Lo_2005} for a review), with a narrow range of SMBH masses. By contrast, molecular gas observations probe a wide range of galaxy and thus SMBH masses, ranging from massive ellipticals \citep[e.g.][]{Cohn_2021, Smith_2021, Ruffa_2023} to dwarfs \citep[e.g.][]{Davis_2020, Nguyen_2022}, constraining SMBH -- galaxy scaling relations across the entire dynamic range.

\subsubsection{Comparing $R_\mathrm{min}/R_\mathrm{Schw}$}
\label{subsec:RSchw}

\begin{figure}
    \centering
    \includegraphics[width=\linewidth]{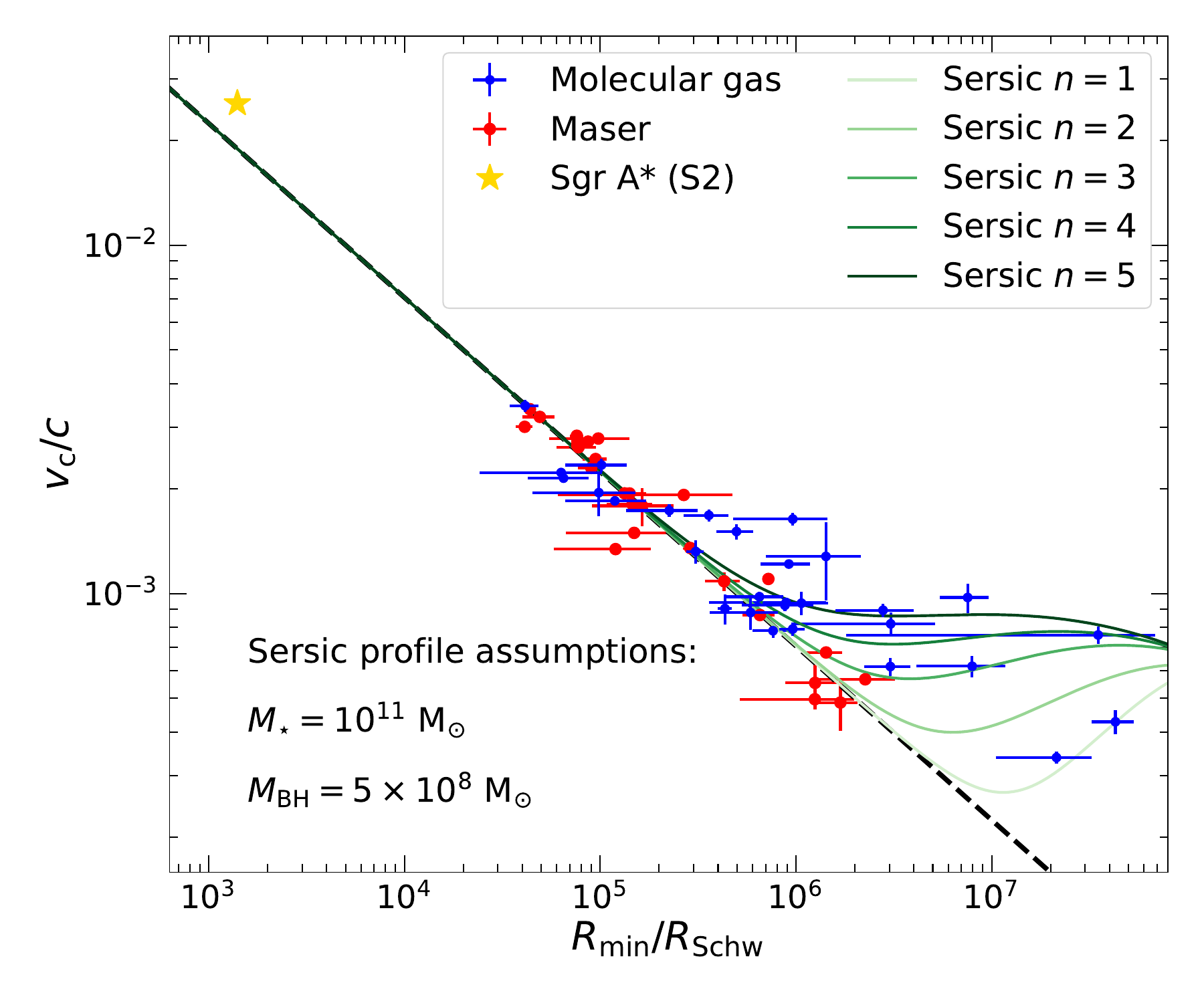}
    \caption{Correlation of the radii and the circular velocities of the innermost kinematic tracer measurements of all galaxies with a SMBH mass measurement using maser or molecular gas kinematics (and the MW), in units of $R_\mathrm{Schw}$ and $c$, respectively. The measurement of Sgr~A$^*$ in the MW uses the peribothron of star S2. The dashed black line shows the Keplerian relation of equation~\eqref{eq:Keplerian_natural_appro}. Observations probing fewer Schwarzschild radii from the SMBH tend to follow the Keplerian curve very well, while observations with lower resolutions (in the unit of $R_\mathrm{Schw}$) deviate from the relation, as the contributions of stars become significant. The expected velocity profiles for a range of S{\'e}rsic indices $n$ are shown for comparison as solid curves in different shades of green, for a galaxy with total stellar mass $M_{*}=10^{11}$~M$_{\odot}$ and SMBH mass $M_\mathrm{BH}=5\times10^8$~M$_{\odot}$. The best molecular gas SMBH mass measurements resolve spatial scales (i.e.\ numbers of Schwarzschild radii) comparable to those of the best maser SMBH mass measurements. In fact, the recent high-resolution WISDOM measurement of NGC~383 (leftmost blue data point; \citealt{Zhang_2024_submitted}) probes material with smaller $R_\mathrm{min}/R_\mathrm{Schw}$ and higher rotational velocities than those of all previous molecular gas {\it and} maser measurements.
    }
    \label{fig:Schw_plot}
\end{figure}

Figure~\ref{fig:Schw_plot} shows the correlation of the radii and the circular velocities of the innermost kinematic tracer measurements of all galaxies with a SMBH mass measurement using maser or molecular gas kinematics (and the MW), in the unit of $R_\mathrm{Schw}$ and $c$, respectively. Because $R_\mathrm{Schw}$ quantifies the physical scales of the motions probed and of processes such as accretion and feedback in the circumnuclear discs, observations with smaller $R_\mathrm{min}/R_\mathrm{Schw}$ trace these processes closer to the SMBHs. Using this metric, the best molecular gas SMBH mass measurements resolve spatial scales (i.e.\ numbers of Schwarzschild radii) comparable to those of the best maser SMBH mass measurements. In fact, the highest-resolution (smallest $R_\mathrm{min}/R_\mathrm{Schw}$) molecular gas measurement is the recent WISDOM measurement of NGC~383 \citep{Zhang_2024_submitted}, that probes material with smaller $R_\mathrm{min}/R_\mathrm{Schw}$ ($\approx41,300$) and higher $V_\mathrm{c}$ ($\approx1040$~\kms) than those of all previous molecular gas {\it and} maser measurements.
This is because, as masers are present in galaxies with SMBH masses typically two orders of magnitude smaller than those of galaxies studied with molecular gas, the poorer angular resolutions of molecular gas observations (compared to those of maser observations) are compensated by the relatively more massive SMBHs probed. Consequently, the highest resolution molecular gas observations can probe physical processes in the circumnuclear discs closer to the SMBHs than maser observations. We note again, however, that this does not guarantee more reliable SMBH mass determinations, as the $R_\mathrm{min}/R_\mathrm{Schw}$ metric does not take into consideration the gravitational contributions of stars.

While observations with small $R_\mathrm{min}/R_\mathrm{Schw}$ mostly follow the Keplerian velocity profile of equation~\eqref{eq:Keplerian_natural_appro}, those with large $R_\mathrm{min}/R_\mathrm{Schw}$ tend to deviate from this relation, because the contributions of the stars to the velocity profiles become significant when $R\sim R_\mathrm{SOI}$. To quantify the typical deviations expected, we generate velocity profiles of a mock galaxy with a total stellar mass $M_{*}=10^{11}$~M$_{\odot}$, a SMBH mass $M_\mathrm{BH}=5\times10^8$~M$_{\odot}$ and different S{\'e}rsic indices $n$. This stellar mass is approximately the midpoint of the total stellar mass dynamic range of the galaxies probed with the molecular gas method, and the $(M_{*},M_\mathrm{BH})$ pair adopted is consistent with the $M_{*}$ -- $M_\mathrm{BH}$ relation of \cite{Bosch_2016}. We note that these parameters are chosen only to illustrate reasonable qualitative trends at large radii, not to reproduce the data points in this region of the diagram.
The mass surface density radial profile of the S{\'e}rsic model is
\begin{equation}
   \Sigma(R)=\Sigma_\mathrm{e}\exp{-b_n\left[\left(\frac{R}{R_\mathrm{e}}\right)^{1/n}-1\right]}\,\,\,.
\end{equation}
We estimate the half-light radius $R_\mathrm{e}$ of each model using the relation of \citet{Shen_2003}:
\begin{equation}
  \frac{R_\mathrm{e}}{\mathrm{kpc}}=
  \begin{dcases}
    0.10\left(\frac{M_{*}}{\mathrm{M}_{\odot}}\right)^{0.14}\left(1+\frac{M_{*}}{3.98\times10^{10}~\mathrm{M}_{\odot}}\right)^{0.25}, & \text{if } n<2.5\\
    2.88\times10^{-6}\left(\frac{M_{*}}{\mathrm{M}_{\odot}}\right)^{0.56}, & \text{if } n>2.5\,\,\,.
    \end{dcases}
\end{equation}
We approximate $b_n$ as
\begin{equation}
b_n\approx2\,n-\frac{1}{3}+\frac{4}{405\,n}+\frac{46}{25515\,n^2}
\end{equation}
\citep[see][]{Ciotti_1999}, where $n$ is the index of the S{\'e}rsic profile. The mass surface density at the effective radius, $\Sigma_\mathrm{e}$, is normalised by the total stellar mass using
\begin{equation}
    M_{*}=\Sigma_\mathrm{e}\left(2\pi R_\mathrm{e}^2\right)\frac{n\,\mathrm{e}^{b_n}}{b_n^{2n}}\Gamma(2n)\,\,\,,
\end{equation}
where $\Gamma$ is the Gamma function. To compute the corresponding three-dimensional mass volume density profiles, we first fit each S{\'e}rsic profile with a set of Gaussians (i.e.\ we perform a MGE fit) using the \texttt{mge\_fit\_1d} procedure of \cite{Cappellari_2002}. We then use the \texttt{mge\_vcirc} procedure of the \textsc{jampy} package \citep{Cappellari_2008,Cappellari_2020} to deproject the Gaussians, add a SMBH at the centre and compute the resulting circular velocity curve.
The resulting curves are shown as solid curves with different shades of green in Figure~\ref{fig:Schw_plot} (with analogous curves in Figures~\ref{fig:SOI_plot} and \ref{fig:Req_plot}). The data points at large $R_\mathrm{min}/R_\mathrm{Schw}$ roughly follow the coloured curves, demonstrating the robustness of our measurements of the innermost tracers. 

\subsubsection{Comparing $R_\mathrm{min}/R_\mathrm{SOI}$}
\label{subsec:RSOI}

\begin{figure}
    \centering
    \includegraphics[width=\linewidth]{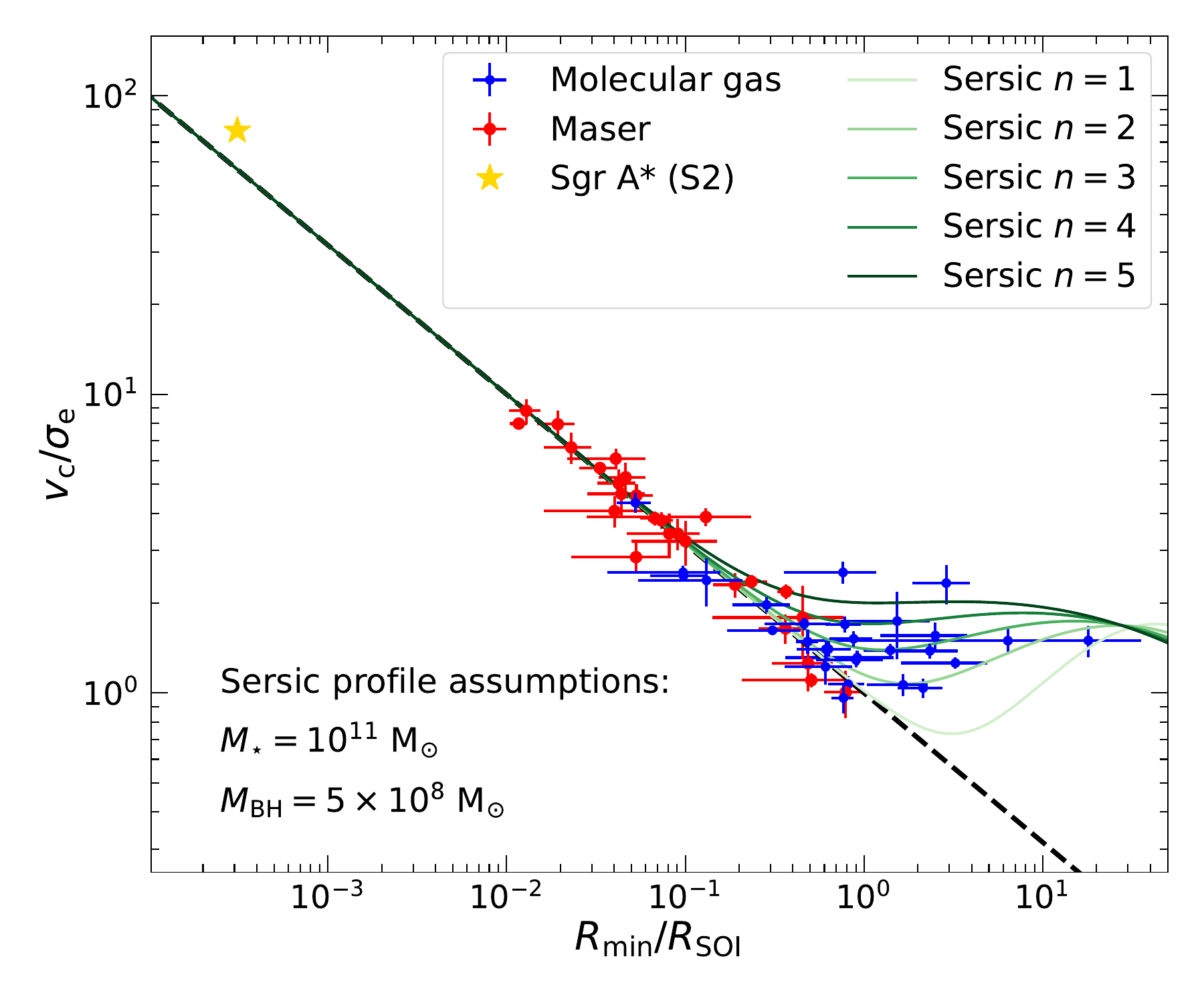}
    \caption{As Figure~\ref{fig:Schw_plot}, but in the unit of $R_\mathrm{SOI}$ and $\sigma_\mathrm{e}$, respectively. 
    The $\sigma_\mathrm{e}$ of each S{\'e}rsic profile is derived by first using the \texttt{jam\_sph\_proj} procedure of the \textsc{jampy} package to compute the second stellar velocity moment ($V_\mathrm{*,RMS}$) map and then calculating the luminosity-weighted sum of $V_\mathrm{*,RMS}^2$ within $R_\mathrm{e}$. Maser observations generally resolve the SOI better than molecular gas observations. However, the recent high-resolution WISDOM measurement of NGC~383 (leftmost blue data point; \citealt{Zhang_2024_submitted}) has a resolution (in the unit of $R_\mathrm{SOI}$) that is in the same range as that of the best maser observations.
    }
    \label{fig:SOI_plot}
\end{figure}

As argued above, detecting material that is fewer Schwarzschild radii away from the SMBH does not in itself guarantee a better SMBH mass measurement, as the measurement precision is more directly related to the factor by which the observations spatially resolve the SOI. Analogously to Figure~\ref{fig:Schw_plot}, Figure~\ref{fig:SOI_plot} thus shows the correlation of the radii and the circular velocities of the innermost kinematic tracer measurements of all galaxies with a SMBH mass measurement using maser or molecular gas kinematics (and the MW), but this time in the unit of $R_\mathrm{SOI}$ and $\sigma_\mathrm{e}$, respectively.

To compute the $\sigma_\mathrm{e}$ of each S{\'e}rsic profile, we adopt the \texttt{jam\_sph\_proj} procedure of the \textsc{jampy} package to calculate a prediction of the second stellar velocity moment ($V_\mathrm{*,RMS}$) from the MGE of the S{\'e}rsic profile. Then, we compute $\sigma_\mathrm{e}^2$ as the luminosity-weighted sum of $V_\mathrm{*,RMS}^2$ within $R_\mathrm{e}$. The resulting $\sigma_\mathrm{e}$ are $110$, $111$, $122$, $126$ and $129$~\kms\ for $n=1$, $2$, $3$, $4$ and $5$, respectively.

Again, almost all observations that resolve the SOI ($R_\mathrm{min}/R_\mathrm{SOI}<1$) follow the Keplerian circular velocity curve, while observations with innermost tracers farther from the SMBHs than $R_\mathrm{SOI}$ are more substantially affected by the galaxies' stellar contents. SMBH mass measurements are nevertheless possible for such observations, if the stellar mass models are sufficiently accurate. In those cases the best-fitting SMBH masses usually have large uncertainties, as the measurements are sensitive to uncertainties in the stellar mass models 
(and unrealistically small uncertainties would almost certainly imply underestimated systematic uncertainties). So far, all SMBH mass measurements using observations with $R>R_\mathrm{SOI}$ are measurements with the molecular gas method, as some of the targeted galaxies were revealed to have central holes in their circumnuclear CO(2-1) or CO(3-2) emission \citep[e.g.][]{Davis_2018, Smith_2019, Smith_2021}. These holes may imply that the inner edges of the molecular gas discs do not reach the SOI or that the molecular gas within the SOI only emits at higher CO transitions and/or only exists as higher-density molecular tracers. In these cases, the radii of the innermost kinematic tracers are greater than $R_\mathrm{SOI}$ regardless of the angular resolutions of the observations.

The maser observations generally resolve the SOI better than the molecular gas observations, in agreement with the generally better SMBH mass precision they offer. Nevertheless, the highest-resolution molecular gas measurement (smallest $R_\mathrm{min}/R_\mathrm{SOI}$), again the recent ALMA measurement of NGC~383 \citep{Zhang_2024_submitted}, has a resolution ($R_\mathrm{min}/R_\mathrm{SOI}=0.05$) that is in the same range as that of the best maser observations. As ALMA's best angular resolution at the frequency of CO(2-1) ($\approx0\farcs018$) is about $40\%$ of $R_\mathrm{min}$ of this measurement, future ALMA molecular gas observations with the most extended configurations can provide comparable $R_\mathrm{min}/R_\mathrm{SOI}$ and thus SMBH mass precision as the highest-resolution measurements using masers (assuming the observations can achieve sufficient signal-to-noise ratios).

Although the angular resolutions of VLBI maser observations are generally much higher than those of ALMA molecular gas observations, the angular sizes of the SOI of galaxies with maser kinematic SMBH measurements are also usually much smaller than those of galaxies with molecular gas kinematic SMBH measurements, partly because of the smaller $M_\mathrm{BH}$ of the galaxies with maser emission, and partly because maser-hosting galaxies are on average farther away due to the scarcity of masers. 
For these reasons, future molecular gas observations with the longest baselines of ALMA can relatively easily achieve the same $R_\mathrm{min}/R_\mathrm{SOI}$ as those of the best maser measurements so far. By contrast, megamaser observations have already exploited the highest angular resolution of current facilities (e.g.\ $\approx0.35$~mas at the frequency of water masers for the Very Long Baseline Array; \citealt{VLBA_1995}). It is thus more challenging to improve the precision of SMBH measurements using masers. 

\subsubsection{Comparing $R_\mathrm{min}/R_\mathrm{eq}$}
\label{subsec:Req}

\begin{figure}
    \centering
    \includegraphics[width=\linewidth]{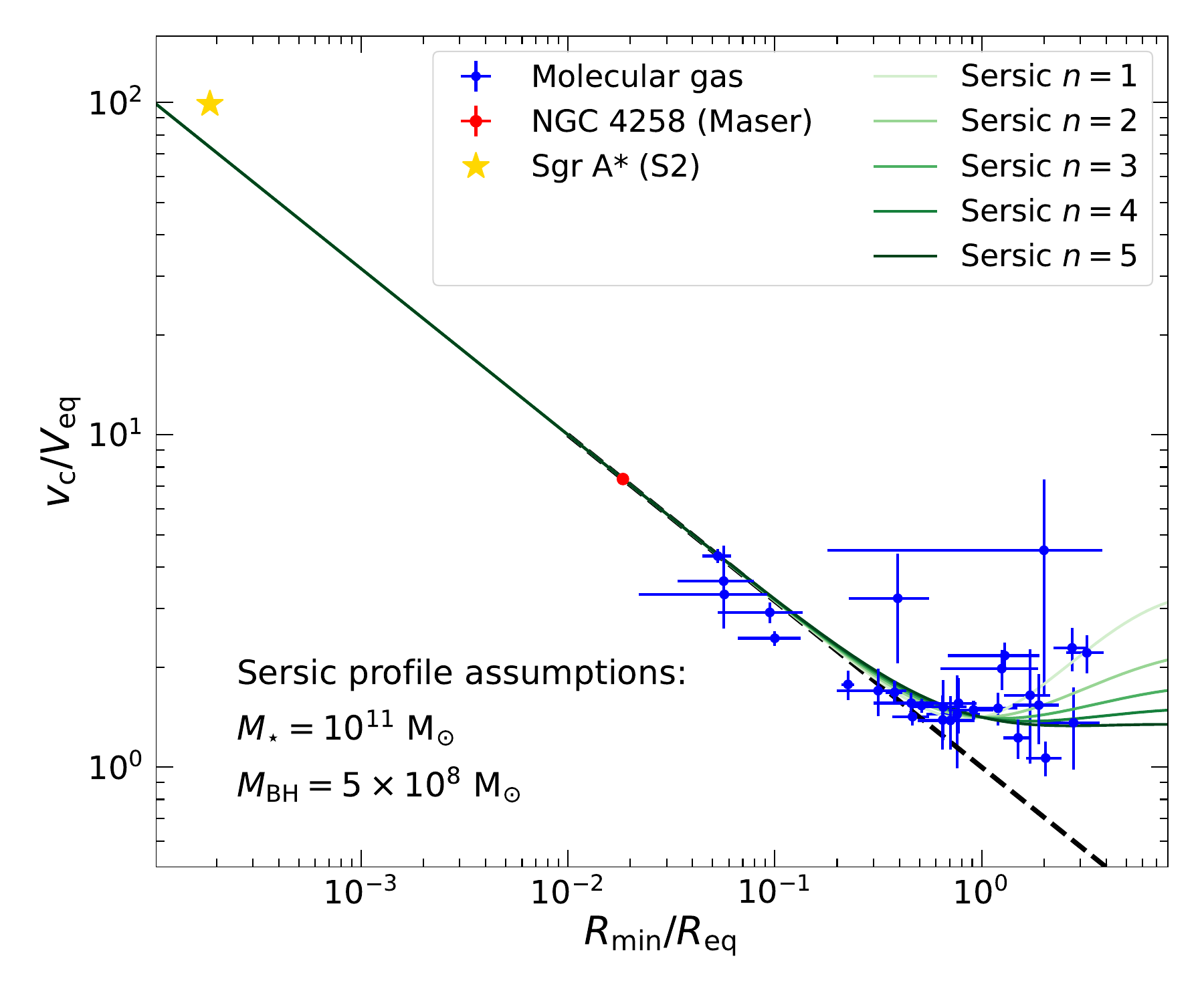}
    \caption{As Figure~\ref{fig:Schw_plot}, but in the unit of $R_\mathrm{eq}$ and $V_\mathrm{eq}$, respectively. The only maser galaxy for which we can compute $R_\mathrm{eq}$ is NGC~4258, which has the largest angular SOI size and the highest resolution SMBH measurement (using either $R_\mathrm{min}/R_\mathrm{Schw}$ or $R_\mathrm{min}/R_\mathrm{SOI}$ as the metric) of all maser galaxies. So far, molecular gas observations have fewer resolution elements across $R_\mathrm{eq}$ than the maser observations of NGC~4258, consistent with the results using $R_\mathrm{SOI}$ as the metric.}
    \label{fig:Req_plot}
\end{figure}

Because $R_\mathrm{SOI}\equiv GM_\mathrm{BH}/\sigma_\mathrm{e}^2$ is only an approximation to the actual SOI radius, we repeat our comparison with $R_\mathrm{SOI}$ replaced by the equality radius $R_\mathrm{eq}$, the formal definition of the SOI radius. Figure~\ref{fig:Req_plot} thus shows the correlation of the radii and the circular velocities of the innermost kinematic tracer measurements of all galaxies with a SMBH mass measurement using maser or molecular gas kinematics (and the MW), this time in the unit of $R_\mathrm{eq}$ and $V_\mathrm{eq}$, respectively. Although only one maser galaxy (NGC 4258) with a SMBH mass determination has an $R_\mathrm{eq}$ measurement, this galaxy also has the smallest $R_\mathrm{min}/R_\mathrm{Schw}$ and $R_\mathrm{min}/R_\mathrm{SOI}$ ratios of all maser galaxies, so the maser data point in Figure~\ref{fig:Req_plot} represents the best maser observations. These observations still have more resolution elements across the SOI than the molecular gas observations (using $R_\mathrm{min}/R_\mathrm{eq}$ as the metric), consistent with the results using the $R_\mathrm{min}/R_\mathrm{SOI}$ metric. However, again, ALMA observations with the most extended configurations can relatively easily produce SMBH measurements with a $R_\mathrm{min}/R_\mathrm{eq}$ and a precision comparable to the highest-resolution measurements using masers.

\subsection{Other considerations}

In practice, achieving a precise SMBH mass using molecular gas observations requires not only high spatial resolution but also a precise determination of the disc inclination, as the SMBH mass depends strongly on the inclination through the deprojection of the velocities: $M_\mathrm{BH}\propto\sin^{-2}i$. Due to this $\sin^{-2}i$ dependence, any inclination uncertainty contributes more to the $M_\mathrm{BH}$ uncertainty budget for smaller (i.e.\ more face-on) inclinations. As maser discs can only be observed nearly edge-on, inclination uncertainties impact SMBH mass uncertainties less for the maser method than for the molecular gas method.

Another advantage of the maser method is that it sometimes enables an independent galaxy distance determination \citep[e.g.][]{Kuo_2013, Pesce_2020}. As $M_\mathrm{BH}$ scales linearly with distance, the distance uncertainty often dominates the $M_\mathrm{BH}$ uncertainty budget, even though it is customary not to include nor quote the distance uncertainty in other $M_\mathrm{BH}$ measurements (as rescaling the SMBH mass to a different distance is straightforward and does not require redoing any fit). An independent distance determination using masers thus improves the true SMBH mass precision (including the distance uncertainty), if the distance obtained is more precise than other distance estimates (obtained using e.g.\ the Tully-Fisher relation; \citealt{Tully_1977}).

On the other hand, a clear advantage of the molecular gas method is that it yields truly three-dimensional data and thus a full two-dimensional velocity map of each circumnuclear disc, while the maser method yields effectively one-dimensional kinematics, i.e.\ the velocities of a few maser spots along the kinematic major axis of the disc only. With many more data points sampling all azimuths, the molecular gas method can provide a reliable SMBH mass even when the data have relatively lower signal-to-noise ratios. Moreover, two-dimensional velocity information allows to properly constrain non-circular motions \citep[e.g.][]{Lelli_2022} and the internal structure of the disc (e.g.\ warps; \citealt{Nguyen_2020, Ruffa_2023}), that can substantially affect the SMBH mass measurement. Modelling these effects is much more difficult using maser data.

Another advantage of the molecular gas method is that molecular gas observations help constrain the mass distribution of the molecular gas disc itself. An accurate molecular gas mass profile allows to disentangle the dynamical effects of the SMBH from those of the self-gravity of the molecular gas disc, thus reducing the systematic uncertainty of the SMBH mass measurement. This is especially crucial when the total molecular gas mass within $R_\mathrm{min}$ is a considerable fraction of the SMBH and/or stellar mass. Measurements with masers, by contrast, require many assumptions to model the disc self-gravity indirectly \citep[e.g.][]{Lodato_2003}.
 
\section{Conclusions}
\label{sec:conclude}

The mass of a SMBH can be measured using spectroscopic observations of kinematic tracers at sufficiently small radii, such that the mass enclosed within the tracer's orbit is dominated by the SMBH. Such tracers then follow a Keplerian circular velocity curve, $V_\mathrm{c}\propto R^{-1/2}$, that can be written in $M_\mathrm{BH}$-independent forms (equations~\ref{eq:Keplerian_natural_appro}, \ref{eq:Keplerian_rsoi_appro} and \ref{eq:Keplerian_req_appro}), affording a fair comparison between SMBH mass measurements that probe different SMBH masses and angular scales. In this paper, we have compared SMBH mass measurements using molecular gas and maser observations, adopting as the metric the radii of their innermost kinematic tracers divided by $R_\mathrm{Schw}$, $R_\mathrm{SOI}$ and $R_\mathrm{eq}$, respectively. We have thus shown that the best molecular gas observations resolve material fewer Schwarzschild radii away from the SMBHs than the best maser observations, so molecular gas observations can probe motions and physical processes closer to the SMBHs than the maser method. Conversely, the best maser observations typically resolve the SMBHs' SOI better than the best molecular gas observations, whether the SOI is defined using the effective stellar velocity dispersion $\sigma_\mathrm{e}$ or the equality radius $R_\mathrm{eq}$. Already, molecular gas observations using the most extended configurations of ALMA can spatially resolve the SOI comparably well, leading to SMBH masses as precise as the most precise SMBH masses derived using masers.

If we accept the claim that masers offer "gold standard" measurements of SMBH masses for $M_\mathrm{BH}\sim10^7$~M$_{\odot}$ (as they resolve the smallest physical scales around the SMBHs), we should consider the molecular gas method capable of producing equally precise measurements across a much wider SMBH mass range. Most of these precise measurements will be towards the high-mass end, where the physical size of the SOI is the largest. Nevertheless, the method can also achieve high precision for a closer, less massive SMBH if a regular and dynamically cold molecular gas disc is present at the centre of the galaxy. Galaxies with lower-mass SMBHs are more likely to have turbulent gas kinematics due to the smaller $V_\mathrm{c}$ and stronger impact of stellar feedback, but regular and circularly rotating CO discs suitable for SMBH mass measurements are also present in some \citep[e.g.][]{Davis_2020}. SMBH measurements with molecular gas are thus highly complementary to existing high-precision measurements with masers. 

In practice, achieving high angular and thus spatial resolutions requires long baselines and, in turn, long integration times ($\approx5$~hr with ALMA). Such observations are not necessarily immediately feasible for a large sample of galaxies. However, the potential for the molecular gas method to steadily increase the number of "gold standard" mass measurements across the entire SMBH mass range is clear. Moreover, it remains to be seen whether the two methods yield consistent masses for the same galaxies, as no galaxy with an existing SMBH mass measurement from maser observations so far meets the selection criteria for a SMBH measurement using molecular gas \citep{Liang_2024}.

\section*{Acknowledgements}
HZ acknowledges support from a Science and Technology Facilities Council (STFC) DPhil studentship under grant ST/X508664/1 and the Balliol College J T Hamilton Scholarship in physics. MB was supported by STFC consolidated grant ‘Astrophysics at Oxford’ ST/K00106X/1 and ST/W000903/1. IR acknowledges support from STFC grant ST/S00033X/1. This research used observations made with the NASA/ESA Hubble Space Telescope and obtained from the Hubble Legacy Archive, which is a collaboration between the Space Telescope Science Institute (STScI/NASA), the Space Telescope European Coordinating Facility (ST-ECF/ESA) and the Canadian Astronomy Data Centre (CADC/NRC/CSA). We acknowledge usage of the HyperLeda database (\url{http://leda.univ-lyon1.fr}).

\section*{Data Availability}
There is no new datum associated with this article.



\bibliographystyle{mnras}
\bibliography{refs} 

\bsp
\label{lastpage}

\appendix
\section{Summary of as yet unpublished SMBH mass measurements}
\label{appendix_summary}
This appendix summarises two submitted but as yet unpublished SMBH mass measurements from the WISDOM project.

\subsection{NGC~383}
The SMBH mass of the nearby lenticular galaxy NGC~383 is measured by \cite{Zhang_2024_submitted} using ALMA observations of the $^{12}$CO(2-1) emission line with a synthesised beam FWHM of $\approx16\times8$~pc$^2$ ($0\farcs051\times0\farcs025$). This angular (and thus spatial) resolution is $\approx4$ times better than that of the previous intermediate-resolution measurement by \cite{North_2019} and it spatially resolves the SOI by a factor of $\approx 23$. The observations yield $V_\mathrm{max}\approx634$~\kms\ (corresponding to a deprojected velocity $V_\mathrm{c}\approx1040$~\kms), $\approx1.8$ times higher than that of \cite{North_2019}, as well as evidence for a mild position angle warp and/or non-circular motions within the central $\approx0\farcs3$. By forward modelling the mass distribution and ALMA data cube, \cite{Zhang_2024_submitted} infer a SMBH mass of $(3.59\pm0.20)\times10^9$~M$_{\odot}$ ($1\sigma$ confidence interval), more precise (5\%) but consistent with ($\approx1.4\sigma$ smaller than) the measurement by \cite{North_2019}. The best-fitting SMBH mass is insensitive to varying models of the central warp and/or non-circular motions. The PVD of the ALMA data, overlaid with that of the best-fitting model, is shown in the online supplementary material.

\subsection{NGC~4751}
The SMBH mass of the early-type galaxy NGC~4751 is measured by \cite{Dominiak_2024_submitted2} using ALMA observations of the $^{12}$CO(3–2) line with an angular resolution of $\approx24$~pc ($0.19$~arcsec). The observations reveal a regularly rotating central molecular gas disc with clear central Keplerian motions. By forward-modelling the molecular gas kinematics and data cube, \cite{Dominiak_2024_submitted2} infer a SMBH mass $M_\mathrm{BH}=3.43^{+0.45}_{-0.44}\times10^9$~M$_\odot$ assuming a constant stellar $M/L$, but a SMBH mass $M_\mathrm{BH}=2.79_{-0.57}^{+0.75}\times10^9$~M$_\odot$ assuming a linearly spatially-varying $M/L$. We adopt the linearly varying $M/L$ model as it agrees more closely with the SMBH mass derived through stellar kinematics in the same paper. The PVD of the ALMA data, overlaid with that of the best-fitting linearly-varying $M/L$ model, is shown in the online supplementary material.

\section{Examples of innermost tracer measurements of molecular gas observations}
\label{appendix_examples}

\begin{figure}
    \centering
    \includegraphics[width=\linewidth]{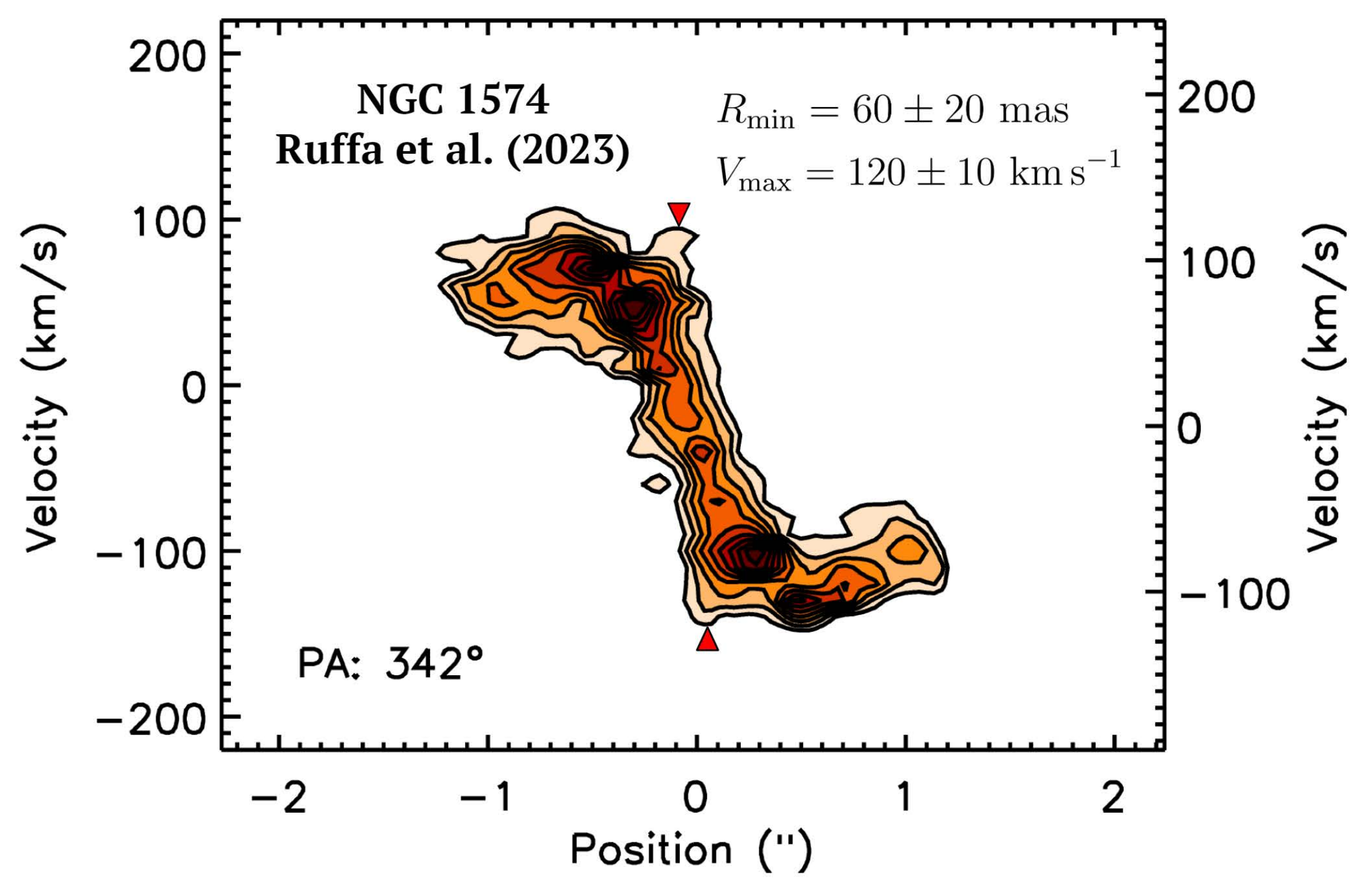}
    \caption{Major-axis position-velocity diagram of the CO(2-1) emission of the galaxy NGC~1574 (Fig.~5d of \citealt{Ruffa_2023}). The red triangles indicate the data points we adopt as the innermost kinematic tracer.}
    \label{fig:NGC1574_PVD}
\end{figure}

\begin{figure}
    \centering
    \includegraphics[width=\linewidth]{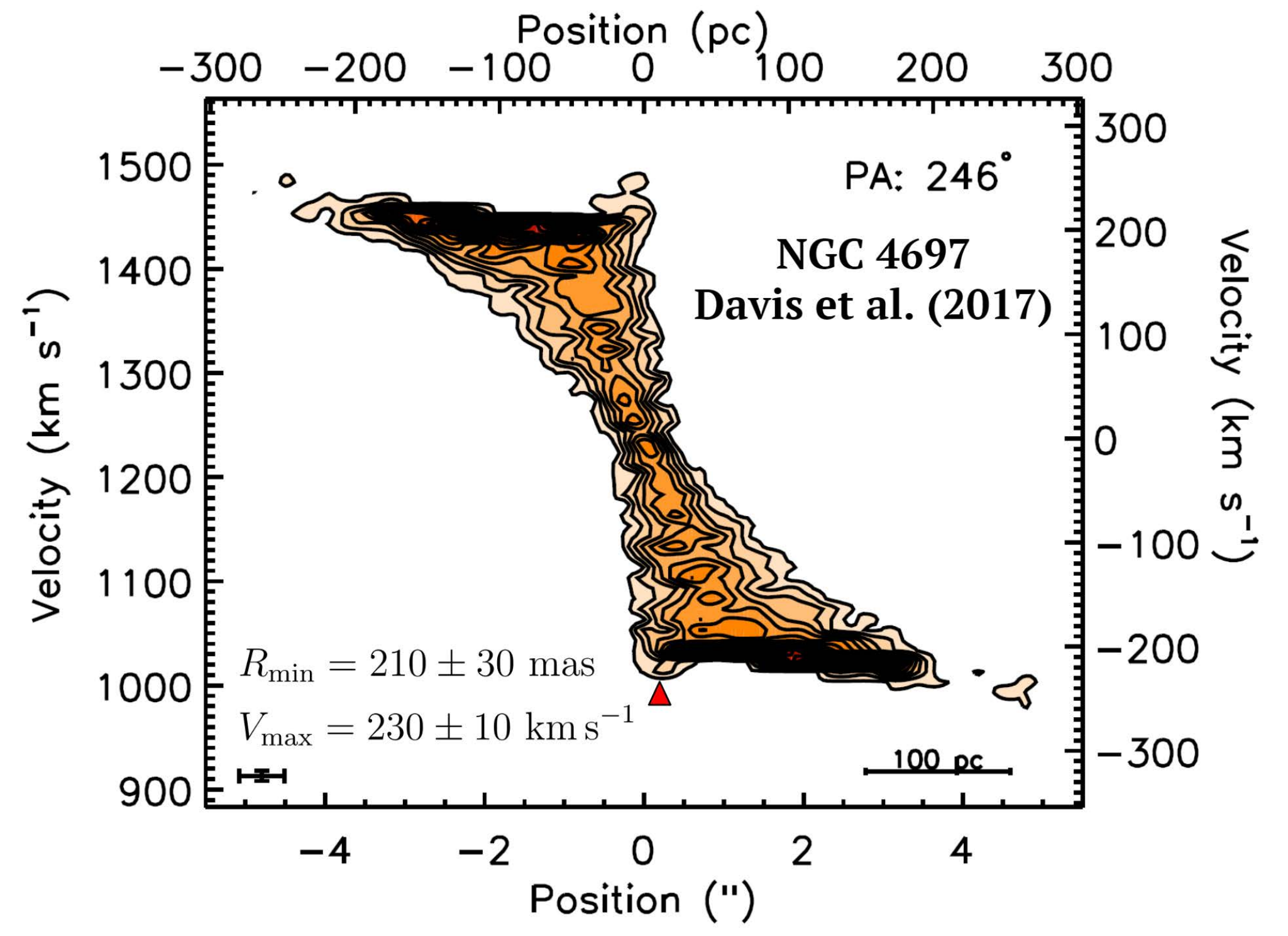}
    \caption{As Figure~\ref{fig:NGC1574_PVD} but for the galaxy NGC~4697 (Fig.~2 of \citealt{Davis_2017}). The emission at the highest positive velocity may well be noise, so we choose not to take our measurement there.}
    \label{fig:NGC4697_PVD}
\end{figure}

\begin{figure}
    \centering
    \includegraphics[width=\linewidth]{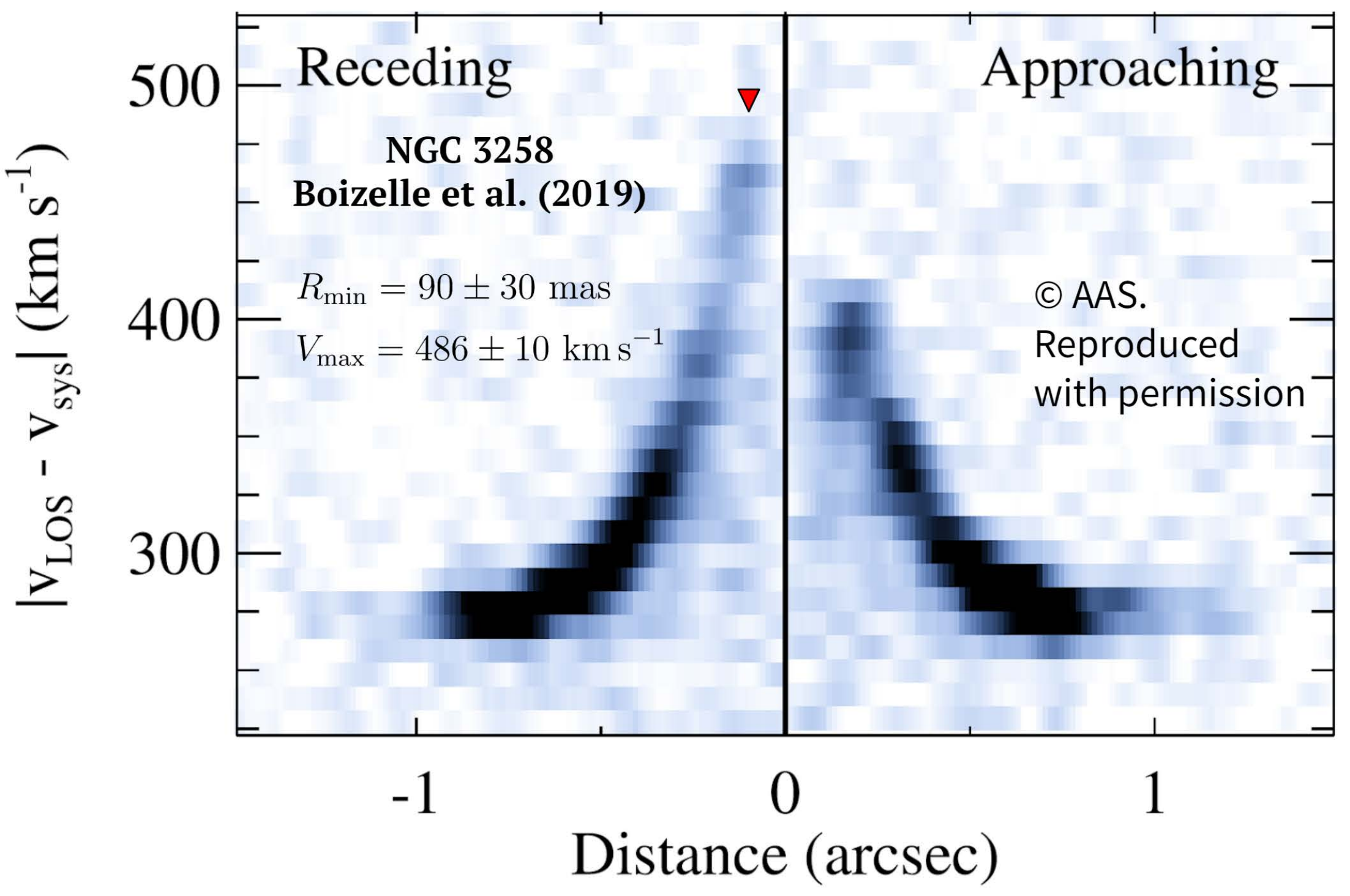}
    \caption{As Figure~\ref{fig:NGC1574_PVD} but for the galaxy NGC~3258 (Fig.~5 of \citealt{Boizelle_2019}). As the redshifted and the blueshifted velocity peaks are strongly asymmetric, we adopt only the higher velocity peak.}
    \label{fig:NGC3258_PVD}
\end{figure}

\begin{figure}
    \centering
    \includegraphics[width=\linewidth]{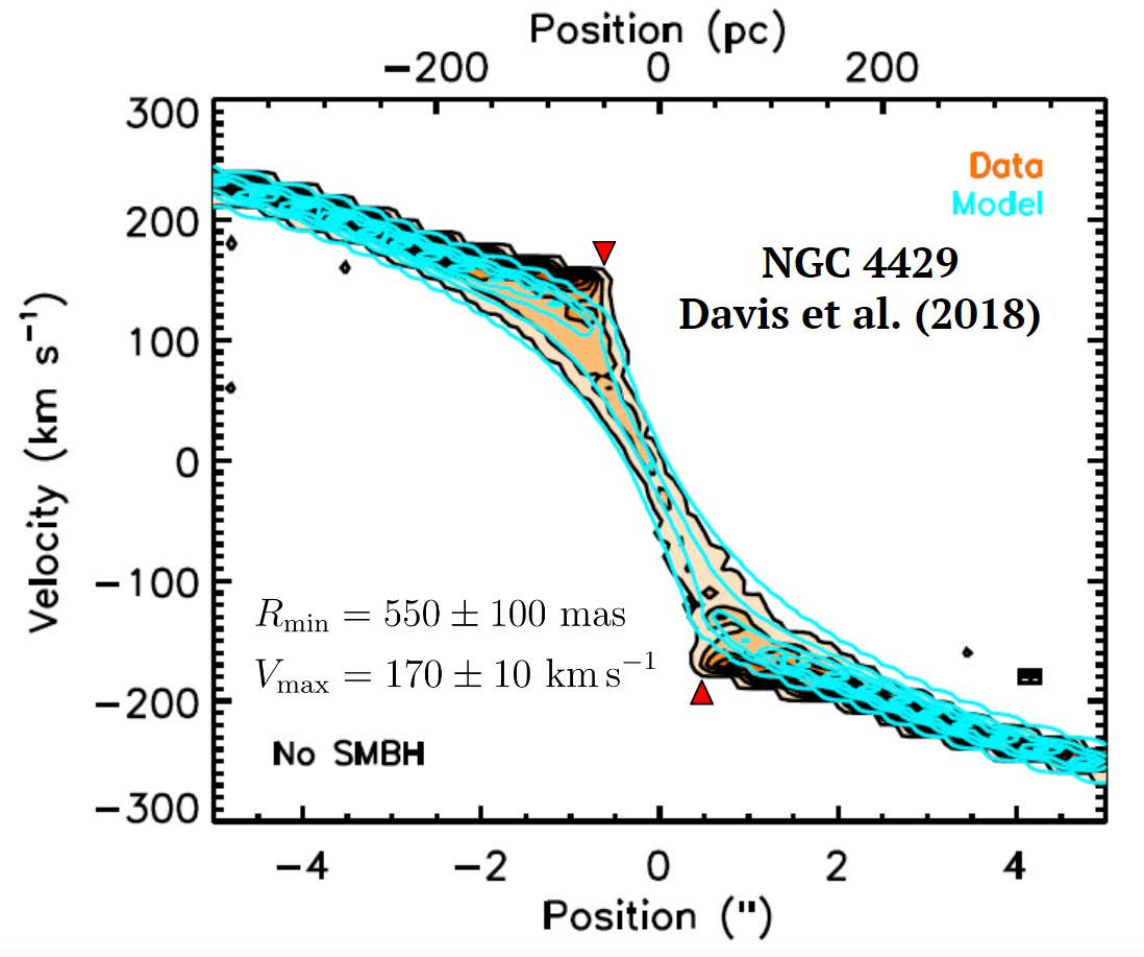}
    \caption{As Figure~\ref{fig:NGC1574_PVD} but for the galaxy NGC~4429 (Fig.~10 of \citealt{Davis_2018}). Although there is no central rise in velocity, we identify the innermost kinematic tracer as the innermost point before the velocity falls off rapidly to zero. The innermost kinematic tracer is outside the SMBH SOI because of a central hole in the CO(2-1) morphology. Yet, a SMBH mass measurement is possible as the velocity of the tracer is higher than that expected from the best-fitting model without a SMBH (cyan contours).}
    \label{fig:NGC4429_PVD}
\end{figure}

\begin{figure}
    \centering
    \includegraphics[width=\linewidth]{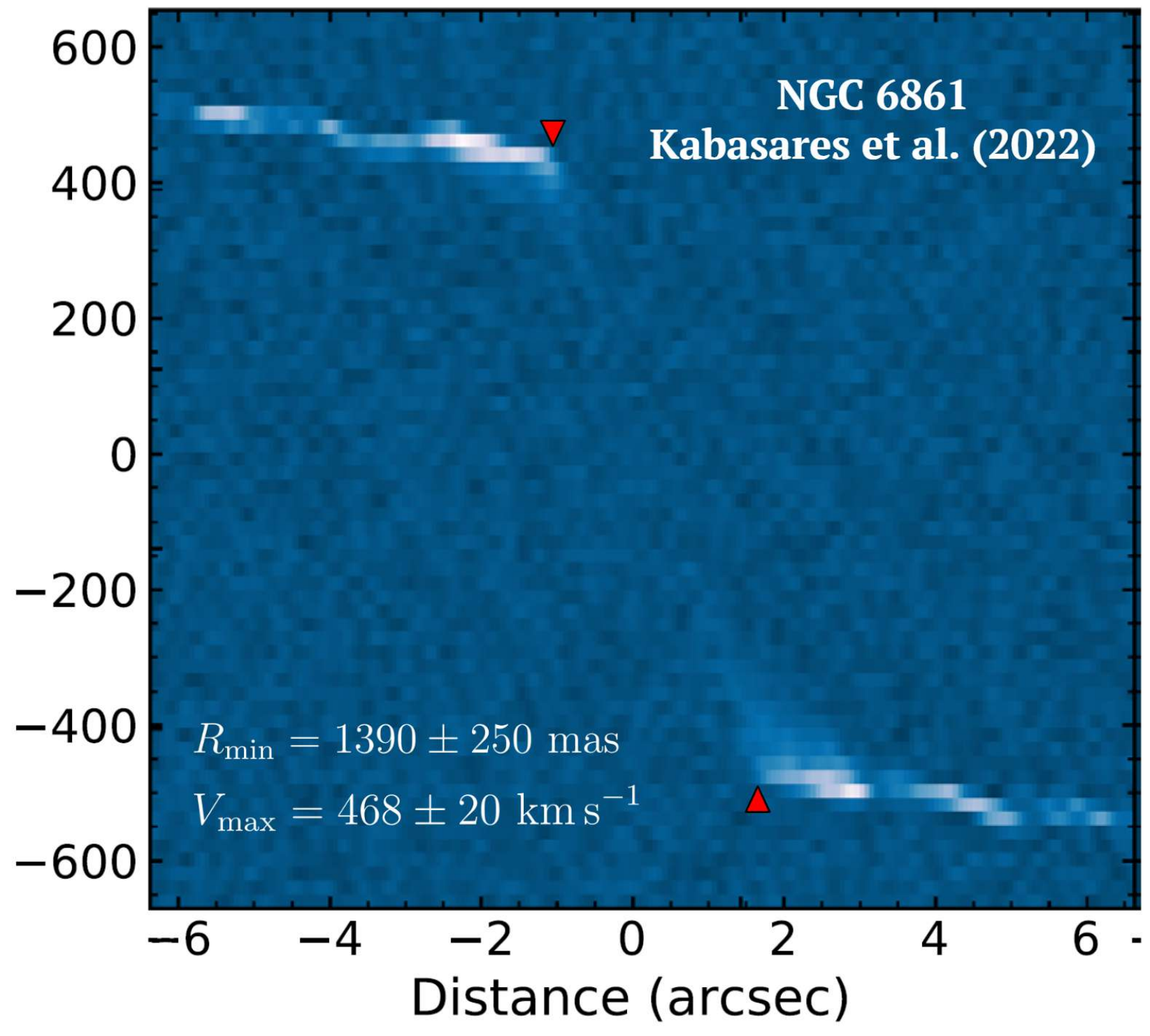}
    \caption{As Figure~\ref{fig:NGC1574_PVD} but for the galaxy NGC~6861 (Fig.~7 of \citealt{Kabasares_2022}). The PVD was plotted using a continuous colour map without contours, so we visually identify the innermost point that is distinguishable from noise.}
    \label{fig:NGC6861_PVD}
\end{figure}

Figures~\ref{fig:NGC1574_PVD} to \ref{fig:NGC6861_PVD} show five examples of our measurements of the innermost kinematic tracers used for SMBH mass measurements with the molecular gas method. The first three examples are from observations that spatially resolve the SOI. The last two examples are from observations that do not detect emitting material within the SOI due to a central hole in the CO(2-1) morphology. Similar figures for all remaining measurements are shown in the online supplementary material.

\end{document}